\documentclass[aip,amsmath,amssymb,reprint]{revtex4-1}

\usepackage{dcolumn}
\usepackage{longtable}
\usepackage{tabularx}
\usepackage{bm}

\usepackage[utf8]{inputenc}
\usepackage[T1]{fontenc}
\usepackage{mathptmx}

\usepackage{amssymb,amsfonts,amsmath}
\usepackage[dvips]{graphicx}
\usepackage[usenames,dvipsnames]{color}
\usepackage{url}
\usepackage[normalem]{ulem}
\usepackage{cancel}
\usepackage[official]{eurosym}

\usepackage{color}
\usepackage{xcolor}

\begin{document}

\preprint{AIP/123-QED}

\title{Optimization with delay-induced bifurcations}

% ------------Define authors and affiliations------------------

\author{Natalia~B.~Janson}
\email[E-mail: ]{N.B.Janson@lboro.ac.uk}
\author{Christopher~J.~Marsden}
\affiliation{Department of Mathematical Sciences, Loughborough University, Loughborough LE11 3TU, United Kingdom}
%\date{\today}
% ----------------Abstract -------------------------------------

\begin{abstract}
  
 Optimization is finding the best solution, which mathematically amounts to locating the global minimum of some cost function. Optimization is traditionally automated with digital or quantum computers, each having their limitations and none guaranteeing an optimal solution. Here we conceive a principle behind optimization based on delay-induced bifurcations, which is potentially implementable in non-quantum analogue devices. Often optimization techniques are interpreted via a particle moving in multi-well energy landscape, and  to prevent  confinement to a  non-global minima they should incorporate mechanisms to overcome barriers between the minima. Namely,
simulated annealing \emph{digitally} emulates pushing a fictitious particle over a barrier by random noise, whereas \emph{quantum} computers utilize tunnelling through barriers.  In our principle, the barriers are effectively destroyed by delay-induced bifurcations.
Although bifurcation scenarios in nonlinear delay differential equations can be very complex and are notoriously difficult to predict, we hypothesize, verify and utilize the finding that they become considerably more predictable in dynamical
systems, where the right-hand side depends only on the delayed variable and represents a gradient of some potential energy function. By tuning the delay introduced into the gradient descent setting, thanks to global bifurcations destroying local attractors, one could force the system to 
spontaneously wander around all minima. This would be similar to noise-induced behavior in simulated annealing, but achieved deterministically. Ideally, a slow increase and then decrease of the delay should automatically push the system towards the global minimum. We explore the possibility of this scenario and formulate some prerequisites.

\end{abstract}


\maketitle

% ---------------- Main part -----------------------------

\begin{quotation}

Can optimization problem be solved without either  relatively slow digital computers, or fast, but costly and difficult to make  quantum computers? Instead of employing complex algorithms or expensive technologies, could we rely on a fast analogue device operating in a semi-automatic manner?
  We conceive and prove a principle   behind optimization, which uses bifurcations caused by time delay in nonlinear   dynamical systems, and could potentially be implemented in analogue circuits. 
  We  hypothesize  and verify that,  in a special  class of delay equations involving the gradient of the cost function, an increase of delay induces a chain of 
  global
 bifurcations, which effectively remove the barriers between the minima  and enable the system to ``explore" the vicinities of all minima.
 Thus, we propose  and demonstrate an alternative way for a candidate optimizer  to overcome the barriers.
 The delay plays the role of  random noise in standard  algorithm-based optimization settings,  such as simulated annealing,  or of  the tunneling effect in quantum computers, and could be promising for obtaining a global minimum. We consider various 
 configurations of the cost function with five minima  with slightly different forms of the barrier-breaking mechanism and demonstrate that in many cases the system converges to the global minimum as desired. 
  Testing with real-life problems and hardware implementation of the proposed principle are the tasks for the future.  Given   that none of the existing optimization 
 approaches
 can fully guarantee the optimal solution,  that digital computers can be too slow, and quantum computers require expensive technologies, our approach
 could represent an attractive alternative  deserving further exploration.

\end{quotation}

\section{Introduction}

\label{sec_intro}

Optimization is a challenging task of finding the best solution out of all the solutions available. Mathematically, optimization amounts to finding a global minimum of some cost (or utility) function that often depends on many variables and possesses many local minima \cite{Horst_handbook_global_optimization_95,Pardalos_handbook_global_optimization_v2_02}. To solve this problem, three main general approaches have been employed based on distinct principles, namely, on an algorithm, an analogue device, and a quantum device.  The arguably most popular approach to date has been based on an \emph{algorithm}, i.e. on a digital computer. 
It assumes step-by-step procedures involving decision-making  dependent on the outcome of the previous step, and usually requires generation of random numbers.  A range of methods allow one to do this with various degrees of accuracy and efficiency \cite{Floudas_global_optimization_review_08,Floudas_encyclopedia_optimization_book08}. 
However, in case of multi-dimensional cost functions with many local minima, computer-based methods can be quite time consuming and do not guarantee an optimal solution within reasonable time. Such limitations inspired exploration of approaches based on potentially faster analogue devices 
\cite{Vichik_analogue_circuit_optimization_CCE14}.  With this, perhaps the most famous and promising idea of a non-digital optimizing device is that of a \emph{quantum} computer \cite{King_D-wave_techreport_2017},  which requires expensive technology.   

 Here we conceive a very different principle behind optimization, which is based on delay-induced bifurcations in nonlinear differential equations. This principle  does not require algorithmic decision-making  at every step and for this reason could potentially be implemented in non-digital, and also non-quantum, analogue devices.  
  We   construct a phenomenological mathematical model  
 of an analogue optimizer of an alternative type,  and prove the concept.

\begin{figure}
\includegraphics[width=0.5\textwidth]{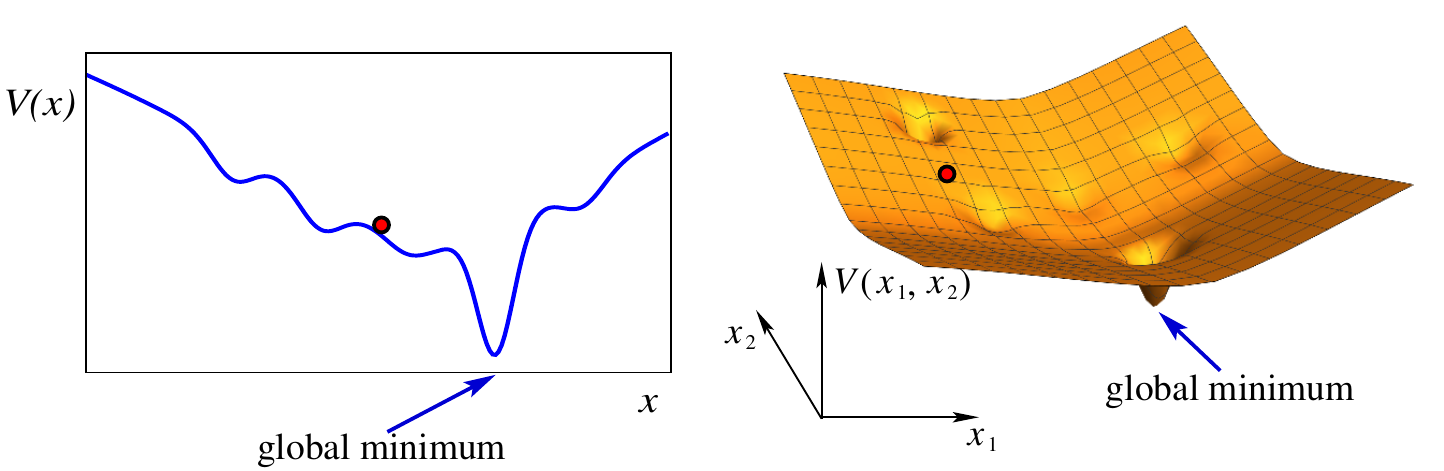}
\caption{Illustration of simple gradient descent. Sample cost functions serving as energy landscapes $V(\boldsymbol{x})$ in Eq.~(\ref{GDS_nodelay}) depending on one (left) and two (right) variables. Red circle represents a zero-mass particle moving in the given  landscape, which automatically converges to the minimum of the potential well where it was put initially.  }
\label{landscape_samples}
\end{figure}

Many popular optimization techniques are based on, or are interpreted via, a gradient descent \cite{Arfken_steepest_descent_book85}, which can be understood as placing a  fictitious zero-mass particle at some randomly chosen position on the energy landscape representing the cost function, and allowing this particle to spontaneously evolve towards the relevant minimum (Fig.~\ref{landscape_samples}). The particle behavior can be described as evolution of a state point  $\boldsymbol{x}$ of a gradient dynamical system (DS)
\begin{equation}
\label{GDS_nodelay}
\dot{\boldsymbol{x}} = \boldsymbol{f}(\boldsymbol{x})=-\nabla V(\boldsymbol{x}), 
\end{equation}
where $\boldsymbol{x}$$=$$(x_1,x_2,\ldots,x_n)$ is a state vector in $\mathbb{R}^n$, 
 which in the context of optimization task becomes the vector of parameters in need of optimization. Also, 
$\dot{\boldsymbol{x}}$ is  the time derivative  of $x$, i.e. $\textrm{d} \boldsymbol{x}/ \textrm{d} t $, $\boldsymbol{f}(\boldsymbol{x})$ is the velocity vector field of the DS,  $V(\boldsymbol{x})$$:$$ \ \mathbb{R}^n \to \mathbb{R}$ plays the role of a scalar energy landscape function at least twice continuously differentiable  \cite{Hirsch_DS_chaos_book04}, and  $\nabla V(\boldsymbol{x}) $$=$$ \left(\frac{\partial{V}}{\partial{x_{1}}}, \ldots, \frac{\partial{V}}{\partial{x_{n}}}\right)$ is its gradient.

However,   the process described by system (\ref{GDS_nodelay}) 
 leads to the local minimum, 
which might not be
 the global one.      In order to find the global minimum starting from arbitrary initial conditions, one needs to introduce into the system a mechanism to overcome the barriers between local minima. Arguably the most popular modifications of the technique above are  
   based on adding stochastic terms in 
Eq.~(\ref{GDS_nodelay}) or in its discrete-time analog \cite{Kushner_stoch_optim_book_78,Kirkpatrick_simulated_annealing_Sci83,Aluffi-Pentini_global_opt_stoch_85, Gelfand_stoch_optim_91}. A distinctive feature of such methods, including one of the most efficient methods called  simulated annealing \cite{Kirkpatrick_simulated_annealing_Sci83,Aluffi-Pentini_global_opt_stoch_85},  is monotonous decrease of the  intensity of the applied random noise (analogue of ``temperature") from some relatively large value to zero, which leads to the convergence  of the solution of the respective differential or difference equation to the global minimum with reasonably large probability. With this, the barriers separating the minima are overcome thanks to random forces pushing the system \emph{over} their tops.   Such approaches principally require algorithmic computation and therefore digital computers. 

Contrary to that, quantum computers utilize quantum annealing, i.e. the ability of a quantum system to \emph{penetrate} the potential barriers between the local minima of their energy function by means of tunnelling, thus dramatically speeding up the search for the global minimum as compared to classical methods. Theoretically, this seems to be the best optimization method, however, its physical implementation 
is technologically highly challenging.  With this, in the existing quantum devices the convergence to the global minimum (ground state) is also achievable with some probability and is therefore not guaranteed \cite{King_D-wave_techreport_2017}.  Generally,  none of the optimization methods available to date are perfect, making exploration of new approaches always valuable.

Here we  propose and explore     an alternative principle enabling the  fictitious particle to overcome the barriers  between the minima 
by inducing global bifurcations in  (\ref{GDS_nodelay}) 
after a 
simple modification. Given that differential equations are implementable in analogue electronic devices, this idea could pave the way for non-digital and non-quantum optimizers.

Namely, we propose to delay the 
 argument
of the right-hand side function in 
(\ref{GDS_nodelay}) by the same amount $\tau$$\ge$$0$ to obtain the following delay-differential equation (DDE)
\begin{equation}
\label{GDS_delay}
\dot{\boldsymbol{x}} = \boldsymbol{f}(\boldsymbol{x}_{\tau}), \quad  \boldsymbol{x}_{\tau}=\boldsymbol{x}(t-\tau), \quad \boldsymbol{f}(\boldsymbol{z})= -\nabla V(\boldsymbol{z}) 
\end{equation}
with $\boldsymbol{x}$, $\boldsymbol{z}$ $\in$ $\mathbb{R}^n$,  assuming that $V(\boldsymbol{z})$$\rightarrow$$\infty$ as $|\boldsymbol{z}|$$\rightarrow$$\infty$. Here, a scalar $\tau$$\ge$$0$ is a single control parameter, whose modification leads to the changes in the available dynamical regimes. 

Inspired by the  theory of bifurcations in ordinary differential equations (ODEs) and by the knowledge of typical delay-induced effects in dynamical systems overviewed in  Section~\ref{sec_theory}, 
we hypothesize that the increase of 
$\tau$ should destroy all local attractors around the minima of  $V$ and create a single large attractor embracing all minima, thus forcing the phase trajectory to spontaneously visit their neighbourhoods. This way, the barriers between the minima would be effectively removed through a fully deterministic mechanism associated with global bifurcations, such as homoclinic bifurcations of fixed points and   of periodic orbits. A useful overview of such homoclinic bifurcations occurring in ODEs can be found in the book from the research group of Vadim Anishchenko \cite{Anishchenko_book_2014}.

Assuming that our hypothesis is true, the desirable optimization procedure would consist of launching the system (\ref{GDS_delay}) from arbitrary initial conditions at $\tau$$=$$0$ and waiting until it reaches one of the local minima, 
then slowly increasing $\tau$ from zero to some positive value, then slowly decreasing $\tau$ to zero, and watching the system converge to the global minimum automatically. This process would be somewhat similar to the simulated annealing in the sense that a control parameter would increase and subsequently decrease to lead the system to the global minimum. However, it would be dissimilar in being deterministic,  which could be advantageous over probabilistic approaches of the simulated and quantum annealings, 
provided that it achieves the same goal. 

 Since a DDE could be implemented in an analogue electronic device, and the increase and  decrease of a single control parameter $\tau$ would be equivalent to turning a handle in an experiment,  the proposed scheme would potentially \emph{not} require sophisticated algorithms and hence a digital computer. However, in this paper we use a digital computer to verify the concept by simulating the relevant equation numerically. 

  Our goal 
is to investigate whether the newly  proposed approach is able to deliver the global minimum successfully, and if so, under what conditions. In Section \ref{sec_theory} we overview the facts about DDEs, which provide the theoretical foundation for our delay-based approach to optimization. In Section \ref{sec_perfomance} we reveal delay-induced bifurcations  and demonstrate how the proposed technique works in systems with five-well landscapes of various configurations  giving rise to various forms of the barrier-breaking mechanism based on global bifurcations. In Section \ref{sec_discussion} we discuss the proposed approach in the context of  quantum optimization and simulated annealing (often used as a benchmark for quantum optimization  \cite{Denchev_quantum_annealing_shape_depenence_PRX16}), and indicate some of its potential advantages and limitations. 

Alongside with exploring the potential applicability of DDEs for optimization, we   reveal some universal behavioral features of nonlinear DDEs of the given class, and thus contribute to the general knowledge of systems with delay.

\section{Theoretical basis for the delay-based optimization:  bifurcations}

 \label{sec_theory} 

The gradient DS of the form (\ref{GDS_nodelay}) can demonstrate only two kinds of evolution: from a certain initial condition it can monotonously tend either to the fixed point located at some local minimum, or to infinity (positive or negative) if the shape of the landscape function $V(\boldsymbol{x})$ permits \cite{Hirsch_DS_chaos_book04}. Importantly, it cannot demonstrate oscillations of any kind, let alone any more complex dynamics.  However, 
it is well known that even if the DS without a delay behaves in a very regular manner, 
introduction of a time delay is likely to change its dynamics dramatically and to induce bifurcations leading to complex oscillations.  Even a single delay term can give birth to periodic \cite{Hale_FDE_book71,Hale_intro_to_FDE_book93}, quasi-periodic and chaotic solutions \cite{Bellen_num_methods_DDE_book03}. This phenomenon occurs thanks to the expansion of the dimension of the phase space of the DS from some finite number without delay to infinity with delay.

There has been a lot of research devoted to the study of various special cases of DSs with delay modelling natural processes \cite{Hutchinson_population_dynamis_delay_ANYAS48,Cunningham_delay_population_PNA54,Mackey_Glass_oscillation_and_chaos_77,Bocharov_delay_immune_response_JTB94,Cooke_two_delays_SEIRS_epidemic_JMB96,Wolkowicz_delay_chemostat_SJAM97,Culshaw_delay_infection_MB00,Smolen_delay_circadian_oscillators_JN01,Nelson_delay_drug_therapy_HIV_MB02,Ursino_delay_respiratory_AJP03,Villasana_delay_tumor_growth_JMB03,Dhamala_delay_neural_network_PRL04}.  In most examples studied, the delay appears in the extra term(s) \emph{added} to the components(s) of the velocity field  $\boldsymbol{f}(\boldsymbol{x})$ of the system under study, like e.g. in \cite{Chow_DDE_added_delay_term_NHMS83,Mallet-Paret_DDE_with_added_delay_term_AMPA86,Chow_DDE_added_delay_term_JDDE89}, or when some terms defining  the velocity field are delayed and some are not \cite{Goodwin_DDE_economics_E51,Cunningham_delay_population_PNA54,Heiden_DDE_chaos_JDE83}.  With this,  the  effects induced by a delay greatly depend on the particular form of the DS under study and on the exact way the delay is introduced.  
It is impossible to predict,  without resorting to numerical bifurcation analysis,  how a \emph{general} non-linear DS would respond if the delay is incorporated into it in some arbitrary manner.

There is a small number of rigorous theoretical results allowing one to qualitatively predict the global behavior of non-linear DDEs for special simple cases of 
  scalar DDEs of the form $\dot{x}$$=$$f(x_{\tau})$ with functions $f$ crossing zero not more than twice.    For $f(z)$$=$$-\frac{\partial V(z)}{\partial z}$,  this would correspond to $V$ having not more than one minimum and/or one maximum. Clearly, such cases do not represent any challenge for optimization and can be handled with the simplest gradient descent (\ref{GDS_nodelay}).

Specifically, in such systems the key phenomena that can be predicted analytically are: (i) the birth of 
 a limit cycle
 localized around the minimum of $V$ \cite{Nussbaum_DDE_periodic_solutions_JDE73,Nussbaum_DDE_periodic_solutions_JDE74}, (ii) a homoclinic bifurcation of a saddle fixed point located at the maximum of $V$ \cite{Walther_DDE_periodic_orbits_from_homoclinic_BCP89,Walther_DDE_Shilnikov_theorem_disser90}, and (iii) a homoclinic bifurcation of a saddle cycle around the maximum of $V$ \cite{Walther_DDE_homoclinic_chaos_from_saddle_orbit_NATMA81,Hale_homoclinic_orbits_DDE_NATMA86}. In \cite{Janson_two-well_delay_2021} it is demonstrated that such homoclinic bifurcations ultimately lead to the disappearance of attractors born from, and localized near, the single minimum of $V$ as the delay $\tau$ grows.  
 
 More realistically, the function $V$ in need of optimization would have more than one minimum, and for such cases \emph{no} rigorous theoretical results predicting homoclinic or other global  bifurcations are available.  
 Striving to predict bifurcations in such systems, we make the following observations. 
 Homoclinic bifurcations are global, i.e. technically non-local, since they do not affect the stability of a saddle point or a cycle, whose manifolds close to form a homoclinic loop. However, the regions of the phase space involved in a homoclinic bifurcation are bounded thanks to the presence of nearby attractors, saddle fixed points or cycles, and/or of their manifolds. So, firstly, homoclinic bifurcations are localized in this sense. Secondly, in (\ref{GDS_delay})
  bifurcations do not change if $V$ is translated in space, which can be easily shown by a trivial variable substitution. Thus, if a single-well $V$ is shifted in the phase space, bifurcations change their locations accordingly, but otherwise stay the same. Thirdly, a smooth multi-well cost function $V$ can be obtained by gluing together, and smoothing out at the gluing points, the segments of   single-well and/or single-hump functions, for which some predictions are possible as described above.

We combine the above observations with theoretical results for single-well and/or single-hump functions $V$, and use essentially induction to hypothesize that in (\ref{GDS_delay}) with a multi-well $V$, the increase of $\tau$ should induce a sequence of global homoclinic-like bifurcations eliminating local attractors one by one. We envision that after all local attractors disappear, at sufficiently large $\tau$ there should exist a single chaotic attractor spanning all maxima and minima of $V$, which would force the phase trajectory to approach the vicinities of all minima, including the global one. This would resemble both simulated annealing, and quantum annealing,  in the sense that the delay would make the imaginary particle overcome the barriers between the local minima and hence act as a substitute for random forces or for quantum tunneling, respectively. 

In \cite{Janson_two-well_delay_2021} global bifurcations were numerically analyzed in 
a scalar version of (\ref{GDS_delay}) with a two-well landscape $V$. 
It was found that the growth of  $\tau$ leads to a chain of  three homoclinic bifurcations, which destroy attractors localized around   two minima of $V$, and to the birth of a large attractor embracing both minima. Although the nature of homoclinic bifurcations varied depending  on the local features of $V$, these were not very important in the sense that, one way or another, they caused the death of   both 
local attractors at sufficiently large $\tau$, and leaving a single large chaos embracing both minima.

 Here we analyze a scalar version of (\ref{GDS_delay}) with a multi-well cost function $V$, which reads
\begin{equation}
\label{dde_whole}
\dot{x}=f(x_{\tau}), \quad f(z)=-\frac{\mathrm{d} V(z)}{\mathrm{d}  z},
\end{equation}
where $x, f, V, z$ $\in$ $\mathbb{R}$, $x_{\tau}$$=$$x(t-\tau)$.   Also,  $V$ is the 
function of a scalar argument, such that $V(z)$$\rightarrow$$\infty$ as $|z|$$\rightarrow$$\infty$. 

In Section \ref{sec_perfomance} we explore and largely confirm the validity of the hypothesis about the chain of  global, homoclinic   or similar, bifurcations evoked by the increasing delay, which continues until all local attractors cease to exist.  
 Note, that for
a landscape with more than two wells, i.e. with more than one saddle fixed point, in addition to homoclinic bifurcations involving the same saddle point, it  becomes reasonable to expect heteroclinic bifurcations involving more than one saddle point. We discuss in what situations the newly proposed procedure can deliver the global minimum of a multi-well landscape.  

In what follows, we use an expression ``instant of bifurcation" for brevity in order to refer to the value of the delay $\tau$ at which the given bifurcation occurs. All bifurcations discussed in this paper occur as $\tau$ \emph{increases} from smaller to larger values. If we consider a fixed point at a value of $\tau$ smaller (or larger) than the value at which a certain bifurcation occurs, we will say that the fixed point is below (or above) the given bifurcation.  
\\

\section{Demonstration of principle}

\label{sec_perfomance}

\begin{table*}[ht]
 \label{tab_1}
\begin{tabular}{|  c |}
\hline 
 {\bf Lowest  minimum is flattest ($x_2^{\textrm{min}}$),  wells have simple shapes. Lowest minimum is found.}  \\ 
 \vbox{ \begin{equation} \label{eq_5well_low5_bro5} V(x)=- \frac{1}{2} \mathrm{e}^{-2(x-1)^2}  - 2 \mathrm{e}^{-(x-5)^2/3}  - \mathrm{e}^{-10(x-7.5)^2/3} 
-  2 \mathrm{e}^{-(x-10)^2/2.9} - \frac{3}{2} \mathrm{e}^{-2(x-12)^2}
 +0.0005 (x-7)^4 - 2.5 \end{equation} } \\
\begin{tabularx}{\textwidth}{ X|X|X|X|X}
 \hline 
 $x_1^{\textrm{min}}$$=$$1.23964$ & $x_2^{\textrm{min}}$$=$$5.01274$  & $x_3^{\textrm{min}}$$=$$7.4972$ & $x_4^{\textrm{min}}$$=$$9.96316$ & $x_5^{\textrm{min}}$$=$$11.8171$ \\  
$V_1^{\textrm{min}}$$=$$-2.41318$ &  $V_2^{\textrm{min}}$$=$$-4.49247$ &  $V_3^{\textrm{min}}$$=$$-3.98079$  &  $V_4^{\textrm{min}}$$=$$-4.46143$  &  $V_5^{\textrm{min}}$$=$$-4.27426$ \\
$J_1^{\textrm{min}}$$=$$- 1.4717$  &  $J_2^{\textrm{min}}$$=$$-1.35261$ & $J_3^{\textrm{min}}$$=$$- 5.61289$ & $J_4^{\textrm{min}}$$=$$- 1.40116$  & $J_5^{\textrm{min}}$$=$$- 4.43609$ \\
$\tau_1^{\textrm{AH1}}$$=$$1.0673$ & $\tau_2^{\textrm{AH1}}$$=$$1.1613$ & $\tau_3^{\textrm{AH1}}$$=$$0.27985$ & 
$\tau_4^{\textrm{AH1}}$$=$$1.1210$ & $\tau_5^{\textrm{AH1}}$$=$$0.35409$ \\
\hline 
  $x_1^{\textrm{max}}$$=$$ 1.82375$ &  $x_2^{\textrm{max}}$$=$$ 6.73759$ &  $x_3^{\textrm{max}}$$=$$ 8.26677$ &  $x_4^{\textrm{max}}$$=$$ 11.1282$ & $\ $ \\
  $J_1^{\textrm{max}}$$=$$ 0.986157$ &  $J_2^{\textrm{max}}$$=$$ 3.47714$ &  $J_3^{\textrm{max}}$$=$$ 3.48976$ &  $J_4^{\textrm{max}}$$=$$ 2.46629$ &  $\ $   \\ 
    $\tau_{1}^{\sigma}$$=$$ 2.0$ &  $\tau_{2}^{\sigma}$$=$$ 0.56799$ &  $\tau_{3}^{\sigma}$$=$$ 0.5659$ &  $\tau_{4}^{\sigma}$$=$$ 0.800$ & $\ $  \\
  $\tau_{1}^{\textrm{AH1}}$$=$$ 4.8166$ & $\tau_2^{\textrm{AH1}}$$=$$ 1.35524$ & $\tau_3^{\textrm{AH1}}$$=$$ 1.35034$ & 
  $\tau_4^{\textrm{AH1}}$$=$$ 1.92596$ &  $\ $  \\
  $\tau_{1}^{\textrm{AH2}}$$=$$ 11.1544$ & $\tau_2^{\textrm{AH2}}$$=$$ 3.1635$ & $\tau_3^{\textrm{AH2}}$$=$$ 3.1508$ & 
  $\tau_4^{\textrm{AH2}}$$=$$ 4.458$ &  $\ $  \\  
 \hline \hline
 \end{tabularx} \\
 {\bf Lowest ($x_2^{\textrm{min}}$) and flattest ($x_4^{\textrm{min}}$) minima are different, wells have simple shapes. Flattest minimum is found. } \\ 
 \vbox{ \begin{equation} \label{eq_5well_low5_bro10} V(x)= - \frac{1}{2} \mathrm{e}^{-2(x-1)^2}  - 2 \mathrm{e}^{-(x-5)^2/3}  - \mathrm{e}^{-10(x-7.5)^2/3} 
-  1.9 \mathrm{e}^{-(x-10)^2/3.5} - 1.3 \mathrm{e}^{-2(x-14)^2}
 +0.0005 (x-7)^4  - 3 \end{equation} } \\
 \begin{tabularx}{\textwidth}{ X|X|X|X|X}
 \hline
 $x_1^{\textrm{min}}$$=$$1.23964$ & $x_2^{\textrm{min}}$$=$$5.01508$  & $x_3^{\textrm{min}}$$=$$7.50702$ & $x_4^{\textrm{min}}$$=$$9.95091$ & $x_5^{\textrm{min}}$$=$$13.859$ \\  
$V_1^{\textrm{min}}$$=$$-2.91318$ &  $V_2^{\textrm{min}}$$=$$-4.99365$ &  $V_3^{\textrm{min}}$$=$$-4.56772$  &  $V_4^{\textrm{min}}$$=$$-4.86134$  &  $V_5^{\textrm{min}}$$=$$-3.16964$ \\
$J_1^{\textrm{min}}$$=$$-1.4717$  &  $J_2^{\textrm{min}}$$=$$-1.34484$ & $J_3^{\textrm{min}}$$=$$-5.67232$ & $J_4^{\textrm{min}}$$=$$-1.12993$  & $J_5^{\textrm{min}}$$=$$-4.76601$ \\
$\tau_1$$=$$1.06733$ & $\tau_2$$=$$1.16802$ & $\tau_3$$=$$0.276923$ & $\tau_4$$=$$1.39017$ & $\tau_5$$=$$ 0.329583$ \\
\hline 
  $x_1^{\textrm{max}}$$=$$ 1.82375$ &  $x_2^{\textrm{max}}$$=$$ 6.72187$ &  $x_3^{\textrm{max}}$$=$$ 8.2824$ &  $x_4^{\textrm{max}}$$=$$ 12.9898$ & $\ $ \\
  $J_1^{\textrm{max}}$$=$$ 0.986157$ &  $J_2^{\textrm{max}}$$=$$ 3.43326$ &  $J_3^{\textrm{max}}$$=$$ 3.20731$ &  $J_4^{\textrm{max}}$$=$$ 2.21328$ &  $\ $   \\ 
    $\tau_{1}^{\sigma}$$=$$ 2.0$ &  $\tau_{2}^{\sigma}$$=$$ 0.575255$ &  $\tau_{3}^{\sigma}$$=$$ 0.61578$ &  
    $\tau_{4}^{\sigma}$$=$$ 0.89234$ & $\ $  \\
  $\tau_{1}^{\textrm{AH1}}$$=$$ 4.77854$ & $\tau_2^{\textrm{AH1}}$$=$$ 1.37257$ & 
  $\tau_3^{\textrm{AH1}}$$=$$ 1.46927$ & 
  $\tau_4^{\textrm{AH1}}$$=$$ 2.12914$ &  $\ $  \\
  $\tau_{1}^{\textrm{AH2}}$$=$$ 11.1499$ & $\tau_2^{\textrm{AH2}}$$=$$ 3.20266$ & $\tau_3^{\textrm{AH2}}$$=$$ 3.42829$ & 
  $\tau_4^{\textrm{AH2}}$$=$$ 4.96799$ &  $\ $  \\  
 \hline \hline
  \end{tabularx} \\
 {\bf  Lowest ($x_5^{\textrm{min}}$), flattest ($x_3^{\textrm{min}}$) and found ($x_4^{\textrm{min}}$) minima are different, wells have    complex shapes. } \\ 
 {\bf  Neither lowest, nor flattest minimum is found.} \\ 
 \vbox{ \begin{equation} \label{eq_5well_chaos_mess} V(x)= 2 \bigg( -\frac{3}{4} \mathrm{e}^{-x^2}  - \frac{1}{4} \mathrm{e}^{-(x-3)^2}  - \frac{1}{4} \mathrm{e}^{-(x-6)^2} 
-  0.2 \mathrm{e}^{-4(x-9)^2/3} -  \frac{2}{5} \mathrm{e}^{-(x-12)^2}  + 0.1 \mathrm{e}^{-10(x-8)^2} +  0.1 \mathrm{e}^{-10(x-5)^2} \bigg)
 +0.0005 (x-7)^4 -2  \end{equation} } \\
 \begin{tabularx}{\textwidth}{ X|X|X|X|X}
 \hline
 $x_1^{\textrm{min}}$$=$$0.218507$ & $x_2^{\textrm{min}}$$=$$3.11875$  & $x_3^{\textrm{min}}$$=$$6.00182$ & $x_4^{\textrm{min}}$$=$$8.98567$ & $x_5^{\textrm{min}}$$=$$11.8539$ \\  
$V_1^{\textrm{min}}$$=$$- 2.37281$ &  $V_2^{\textrm{min}}$$=$$- 2.37975$ &  $V_3^{\textrm{min}}$$=$$-2.49956$  &  $V_4^{\textrm{min}}$$=$$-2.39226$  &  $V_5^{\textrm{min}}$$=$$-2.50557$ \\
$J_1^{\textrm{min}}$$=$$- 2.85663$  &  $J_2^{\textrm{min}}$$=$$- 1.0414$ & $J_3^{\textrm{min}}$$=$$- 1.00708$ & $J_4^{\textrm{min}}$$=$$- 1.08852$  & $J_5^{\textrm{min}}$$=$$- 1.64032$ \\
$\tau_{1}^{\textrm{AH1}}$$=$$0.54987$ & $\tau_2^{\textrm{AH1}}$$=$$1.50835$ & $\tau_3^{\textrm{AH1}}$$=$$1.55975$ & $\tau_4^{\textrm{AH1}}$$=$$1.443056$ & $\tau_5^{\textrm{AH1}}$$=$$0.9576$ \\
\hline 
  $x_1^{\textrm{max}}$$=$$ 1.48726$ &  $x_2^{\textrm{max}}$$=$$ 4.91839$ &  $x_3^{\textrm{max}}$$=$$ 7.946$ &  $x_4^{\textrm{max}}$$=$$ 10.3872$ & $\ $ \\
  $J_1^{\textrm{max}}$$=$$ 1.30507$ &  $J_2^{\textrm{max}}$$=$$ 3.79408$ &  $J_3^{\textrm{max}}$$=$$4.27797$ &  $J_4^{\textrm{max}}$$=$$ 0.768708$ &  $\ $   \\ 
    $\tau_{1}^{\sigma}$$=$$ 1.5133$ &  $\tau_{2}^{\sigma}$$=$$ 0.52054$ &  $\tau_{3}^{\sigma}$$=$$ 0.46166$ &  
    $\tau_{4}^{\sigma}$$=$$ 2.56925$ & $\ $  \\
  $\tau_{1}^{\textrm{AH1}}$$=$$ 3.61084$ & $\tau_2^{\textrm{AH1}}$$=$$ 1.24204$ & 
  $\tau_3^{\textrm{AH1}}$$=$$ 1.10155$ &   $\tau_4^{\textrm{AH1}}$$=$$ 6.13027$ &  $\ $  \\
  $\tau_{1}^{\textrm{AH2}}$$=$$ 8.42529$ & $\tau_2^{\textrm{AH2}}$$=$$ 2.89809$ & 
  $\tau_3^{\textrm{AH2}}$$=$$ 2.57028$ &   $\tau_4^{\textrm{AH2}}$$=$$ 14.304$ &  $\ $  \\  
 \hline \hline
 \end{tabularx} \\
\end{tabular}
\caption{Five-well landscapes $V(x)$ for Eq.~(\ref{dde_whole}) with various local properties used to illustrate the performance of delay-based approach to optimization, shown by blue lines in Figs.~\ref{fig_5_well_low5_broad5_bd}(a), \ref{fig_5_well_low5_broad10_bd}(a) and \ref{fig_5_well_chaos_mess_bd}(a). Different constant terms were added to $V(x)$ for a more convenient graphical representation of various functions these figures. Below each expression for $V$, key features of their local minima $x_i^{\textrm{min}}$ and maxima $x_j^{\textrm{max}}$ are given, with $i=1,\ldots,5$ and $j=1,\ldots,4$. Namely, $V_i^{\textrm{min}}$ are the depths of the minima, $J_i^{\textrm{min}}$/$J_j^{\textrm{max}}$  are the values of the Jacobian $J$$=$$f'$$=$$V''$ at the respective minima/maxima, $\tau_{i}^{\textrm{AH1}}$ are the values of $\tau$ at which the relevant fixed points udergo the first Andronov-Hopf (AH) bifurcation. For the maxima, $\tau_{j}^{\sigma}$ are the values of $\tau$ at which the saddle quantity switches from negative to positive, and $\tau_{j}^{\textrm{AH2}}$ are the values of $\tau$ at which the fixed points undergo the second AH bifurcation. }
\end{table*}

In the given section we demonstrate the performance of our delay-based 
 approach
using three five-well landscapes with subtly different local properties leading to noticeably different delay-induced effects,  and specifically to different orders and forms of bifurcation mechanisms in which the barriers between the local minima are removed. Namely, the first two landscapes considered in Subsections \ref{sec_global} and \ref{sec_flattest} have potential wells of a relatively simple shape, such that the respective functions $f$ of (\ref{dde_whole}) do not oscillate between any two consecutive zero-crossings (see red lines in Figs.~\ref{fig_5_well_low5_broad5_bd}(a) and \ref{fig_5_well_low5_broad10_bd}(a)).  In the third example analysed in Subsection \ref{sec_arbitrary}, some of the wells have more complex shapes, such that within them the function $f$ displays oscillations, as shown by red line in Fig.~\ref{fig_5_well_chaos_mess_bd}(a) between $x$$=$$3$ and $x$$=$$10$. 
With this, in the first example the lowest minimum is also the flattest one, whereas in the second and third examples the lowest minima are not the flattest.  We will show that the proposed approach is likely to deliver the global minimum where this minimum is the flattest and the respective potential well is sufficiently wide, and the cost function is of relatively simple shape, i.e. the right-hand side $f$ of the DDE in  (\ref{dde_whole}) does not oscillate between consecutive zero-crossings. \\

\subsection{Reaching the global minimum}
\label{sec_global}

\begin{figure}
\includegraphics[width=0.4\textwidth]{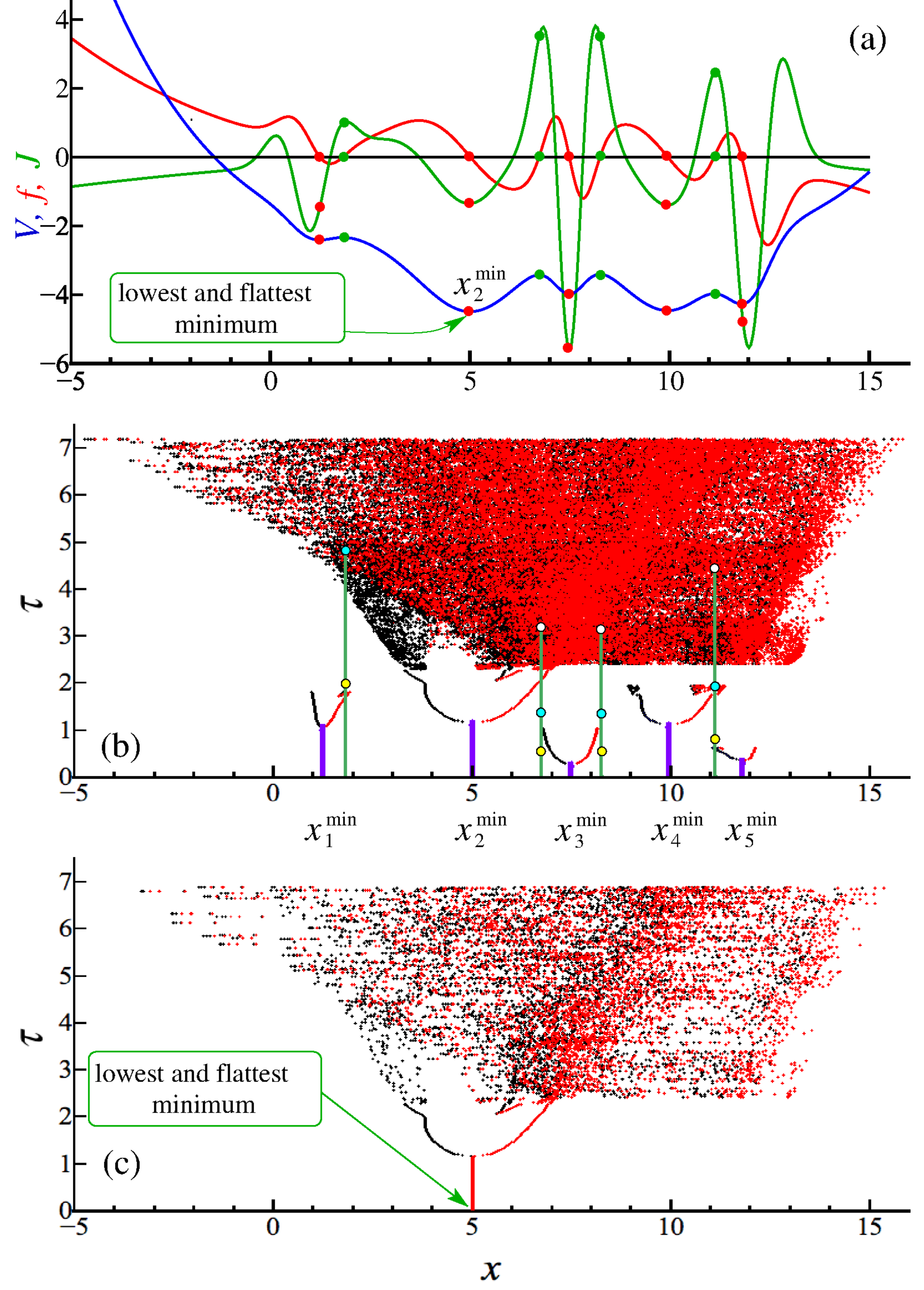}
\caption{Illustration of the case where the lowest minimum $x_2^{\textrm{min}}$ of $V$ is also the flattest. 
{\bf  (a)} Landscape $V(x)$ given  by (\ref{eq_5well_low5_bro5}) (blue line), the function $f(x)$$=$$V'(x)$ of (\ref{dde_whole})  (red line) and its derivative $J(x)$$=$$f'(x)$ (green line). Red/green circles show (i) on blue line, the positions of minima/maxima of $V$, (ii) on red line, the locations of respective fixed points, (iii) on green line, the values of Jacobian $J$ at these fixed points. 
 {\bf (b)} Bifurcation diagram of (\ref{dde_whole})  with landscape (\ref{eq_5well_low5_bro5}) as $\tau$ grows. Black/red dots show local minima/maxima of oscillatory attractors; lilac/green vertical lines show fixed points at the local minima/maxima of $V$; yellow filled circles on green lines  are placed at the values of $\tau$ 
 at which the saddle quantities $\sigma$ of the saddle-foci take zero values; cyan/white filled circles on green lines show first/second Andronov-Hopf bifurcations of the fixed points at the maxima of $V$. 
 {\bf (c) } Behavior of (\ref{dde_whole})   with $V$ given by   (\ref{eq_5well_low5_bro5}) as $\tau$ {\it decreases} slowly from $\tau$$=$$7$ starting from a point on the large chaotic attractor.  Local minima (black dots) and maxima (red dots) of the solution are shown.  The system automatically converges to the global minimum $x_2^{\textrm{min}}$ (see (a) and Table I under (\ref{eq_5well_low5_bro5})). }
\label{fig_5_well_low5_broad5_bd}
\end{figure}

Here we consider the landscape $V$  specified by Eq.~(\ref{eq_5well_low5_bro5}) from Table~I and shown by blue line in Fig.~\ref{fig_5_well_low5_broad5_bd}(a). In the given function $V(x)$, there are five minima marked by red filled circles, and four maxima marked by green filled circles. In what follows, we will denote the minima as $x_i^{\textrm{min}}$ and the maxima as 
$x_j^{\textrm{max}}$, with indices $i$ and $j$ being the numbers of the respective minimum/maximum if counted from the left, so that  $i=1,\ldots,5$ and $j=1,\ldots,4$. 
All minima $x_i^{\textrm{min}}$ together with their depths $V_i^{\textrm{min}}$ are given in Table~I under Eq.~(\ref{eq_5well_low5_bro5}). 
Although the global minimum here is $x_2^{\textrm{min}}$, there is another minimum $x_4^{\textrm{min}}$, which is only marginally higher and would be an almost equivalent choice for the best solution. 
The landscape $V$ has a relatively simple shape, so that $f$$=$$-V'$ does not oscillate between any two consecutive zero-crossings, see red line in Fig.~\ref{fig_5_well_low5_broad5_bd}(a). Below we will first reveal bifurcations occurring in this system as $\tau$ grows, and then the behavior of the system as $\tau$ slowly decreases from a large value. 

\begin{figure*}
\includegraphics[width=0.8\textwidth]{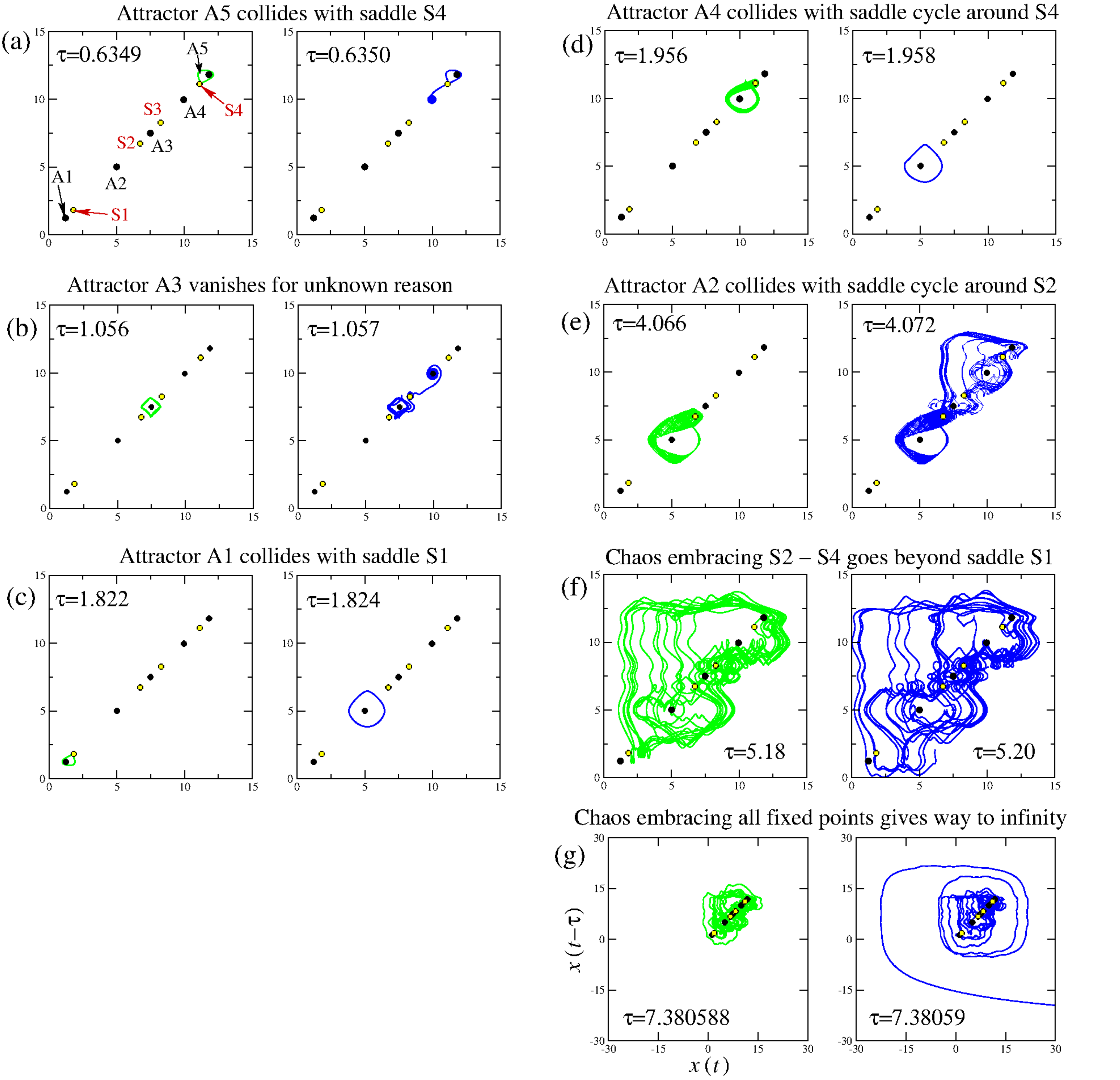}
\caption{Phase portraits on the plane $($$x(t)$, $x(t-\tau)$$)$ illustrating a sequence of non-local bifurcations in (\ref{dde_whole})  with landscape (\ref{eq_5well_low5_bro5}) as $\tau$ increases. 
Here, the lowest minimum $x_2^{\textrm{min}}$ of $V$ is also the flattest. Fixed points at minima/maxima of $V$ are given by black/yellow circles. Panel (a)  shows 
labels of attractors A1 -- A5 at or around the minima, and of saddle fixed points S1 -- S4 at the maxima,   as they are referred to in text. 
The value of  $\tau$ increases from  (a) to (g).  Each panel illustrates the situation just before  (left) and immediately after (right) a certain non-local bifurcation  with the values of $\tau$ given in the fields of each plot. \\
{\bf (a)} Before (left) and after (right) the safe homoclinic loop of the saddle point at the maximum S4. Green line: limit cycle A5 before bifurcation. Blue line: phase trajectory starting near  A5 and converging to fixed point A4  after bifurcation. \\
{\bf (b)} Before (left) and after (right) the global bifurcation eliminating attractor  A3. Green line: chaotic attractor   A3 before bifurcation. Blue line: phase trajectory starting near  A3 and converging to the minimum A4  after bifurcation. \\
{\bf (c)} Before (left) and after (right) the safe homoclinic loop of the saddle point at the maximum S1. Green line: limit cycle  A1 before bifurcation. Blue line: limit cycle A2 to which trajectory converges  after bifurcation starting from the vicinity of  A1. \\
{\bf (d)} Before (left) and after (right) the homoclinic ``loop" formed by the manifolds of the saddle cycle around the maximum S4. Green line: chaotic attractor   A4 before bifurcation. Blue line: limit cycle A2 attracting the trajectory  after bifurcation  from the vicinity of  A4. \\
{\bf (e)} Before (left) and after (right) the homoclinic ``loop" formed by the manifolds of the saddle cycle around the maximum S2. Green line: chaotic attractor  A2 before bifurcation. Blue line: non-localized chaotic attractor born after the bifurcation. \\
{\bf (f)} Before (left) and after (right) the global bifurcation giving birth to large chaos embracing all fixed points. Green line: smaller chaotic attractor  before bifurcation. Blue line: large chaotic attractor born from the bifurcation. \\
{\bf (g)} Death of the last attractor. Green line: large chaos before bifurcation. Blue line: trajectory going to infinity after bifurcation.  
}
\label{fig_5_well_low5_broad5_pp}
\end{figure*}

At any $\tau$,  the equation (\ref{dde_whole}) with  $V$ specified by (\ref{eq_5well_low5_bro5}) has nine fixed points indicated by circles in Fig.~\ref{fig_5_well_low5_broad5_pp}(a). At $\tau$$=$$0$, there are five stable and four unstable fixed points,  both of a node type, which means that the solution $x(t)$ approaches the former and departs the latter in a non-oscillatory manner. At $\tau$$>$$0$, every fixed point has infinitely many pairs of complex-conjugate eigenvalues.  An overview of the theoretical results related to the stability analysis of fixed points in DDEs using Lambert function is available  in
\cite{Asl_DDE_stability_eigenvalues_ASME03}. 
The values of $\tau$ at which local bifurcations of the fixed points occur are determined by the values of the Jacobian $J$$=$$f'$ at these points.    A $k$th eigenvalue $\lambda_k$ of a fixed point of (\ref{dde_whole})
 can be expressed in terms of the Lambert function $W(y)$ with $y$$\in$$\mathbb{R}$ and $W$$\in$$\mathbb{C}$ as
\begin{equation*}
\label{eig_lambert}
\lambda_k = \frac{W_{k}(J \tau)}{\tau}=\frac{JW_{k}(z)}{z}, \quad k=0, \pm 1, \pm 2, \ldots
\end{equation*}
where $J$$=$$f'(x^*)$ and $x^*$ is the fixed point. Note, that the function $W(z)$ has countably many branches. Illustrations of $W(z)$ and of the respective eigenvalues as functions of $\tau$  can be found e.g. in    \cite{Janson_two-well_delay_2021}.

The function $f'$ is shown in Fig.~\ref{fig_5_well_low5_broad5_bd}(a) by a green line, and its values at the fixed points are indicated by red or green filled circles for the minima and maxima, respectively. 

Firstly, consider local bifurcations of the fixed points at the landscape {\it minima} $x_i^{\textrm{min}}$, at which  $J$$<$$0$. The values
$J_i^{\textrm{min}}$ of $J$ at all minima are given in Table~I under Eq.~(\ref{eq_5well_low5_bro5}). 
For  $\tau$$\in$$\left( 0, \frac{1}{e|J_i^{\textrm{min}}|} \right)$ a fixed point $x_i^{\textrm{min}}$ has one real negative  eigenvalue and infinitely many pairs of complex-conjugate eigenvalues with very large negative real parts.  
The latter implies that although $x_i^{\textrm{min}}$ is technically a stable focus, the phase trajectories approach this point in a non-oscillatory manner.  For  $\tau$$\in$$\left[\frac{1}{e|J_i^{\textrm{min}}|}, \frac{\pi}{2|J_i^{\textrm{min}}|} \right)$ the same point has infinitely many pairs of complex-conjugate eigenvalues with negative real parts.  It remains a stable focus, but here the real parts of the leading pair of eigenvalues  are relatively small negative numbers, so the solution near this point oscillates. 

At a  value of $\tau$$=$$\frac{\pi}{2 |J_i^{\textrm{min}}|}$ from this point a stable limit cycle is born via the first Andronov-Hopf (AH) bifurcation. 
The  respective values $\tau_{i}^{\textrm{AH1}}$  of $\tau$ for all minima are given in Table~I below Eq.~(\ref{eq_5well_low5_bro5}). In the bifurcation diagram given in Fig.~\ref{fig_5_well_low5_broad5_bd}(b), the fixed points at the minima $x_i^{\textrm{min}}$ are shown as lilac vertical lines until the values $\tau$$=$$\tau_{i}^{\textrm{AH1}}$, 
i.e. as long as they remain stable. Above these values of $\tau$ they continue to exist as unstable fixed points of a saddle-focus type and are not shown. At even higher values of $\tau$ they undergo more AH bifurcations, which do not restore their stability and instead make them more and more unstable. 

 Also, in Fig.~\ref{fig_5_well_low5_broad5_bd}(b) above every point of the first AH bifurcation of $x_i^{\textrm{min}}$, i.e. above the top of the lilac line, we show both the maxima (red dots) and the minima (black dots) of the newly born stable limit cycles. This style of presenting a bifurcation diagram is  different from a conventional one, in which only one kind of an attractor extremum is usually depicted. Here we need to monitor both extrema in order to observe how a limit cycle grows in size with $\tau$ and eventually approaches a nearby saddle fixed point or another saddle object before vanishing in a homoclinic bifurcation, bearing in mind that the respective saddle object  can exist to any side of the limit cycle.  At higher values of $\tau$, the limit cycles are replaced by non-periodic attractors, for which we continue to show all the local maxima and minima as they can also experience  homoclinic bifurcations. 
 
 Note that at the same values of $\tau$ there can be more than one co-existing non-fixed point attractor. However, in Fig.~\ref{fig_5_well_low5_broad5_bd}(b)  we use the same two colors black and red to depict all such attractors at the given value of $\tau$. The reason is that technically it becomes quite difficult to label different attractors in the presence of so many non-local bifurcations. 
 Therefore, the given bifurcation diagram alone does not allow one to determine either the total number of coexisting attractors, or the span of individual attractors at every value of $\tau$. However, a better understanding of the sequence of bifurcations can be achieved by comparing this diagram with phase portraits  in Fig.~\ref{fig_5_well_low5_broad5_pp} described below. 

Now, consider local bifurcations of the fixed points at the landscape {\it maxima} $x_j^{\textrm{max}}$ indicated by vertical green lines in Fig.~\ref{fig_5_well_low5_broad5_bd}(b), at which  $J$$>$$0$. 
From being unstable nodes at $\tau$$=$$0$, at any positive $\tau$ they turn into saddle-foci  because they have a finite number of pairs of complex-conjugate eigenvalues with positive real parts, and a countably infinite number of complex-conjugate eigenvalue pairs with negative real parts. 
With this, they can also undergo AH bifurcations. Namely, the first AH bifurcation of a saddle-focus fixed point $x_j^{\textrm{max}}$  occurs at $\tau$$=$$\tau_{j}^{\textrm{AH1}}$$=$$\frac{3\pi}{2 J_j^{\textrm{max}}}$, where $J_j^{\textrm{max}}$$>$$0$ is the Jacobian $J$ at this  point. The values of 
$J_j^{\textrm{max}}$ and the respective $\tau_{j}^{\textrm{AH1}}$ for all maxima are given in Table~I under Eq.~(\ref{eq_5well_low5_bro5}).
The first AH bifurcation can give rise to a saddle cycle. All four unstable fixed points $x_j^{\textrm{max}}$ are shown in Fig.~\ref{fig_5_well_low5_broad5_bd}(b) by green vertical lines for the ranges of $\tau$ until their respective second AH bifurcations at 
 $\tau_{j}^{\textrm{AH2}}$$=$$\frac{7\pi}{2 J_j^{\textrm{max}}}$ occurs. An exception is  $x_1^{\textrm{max}}$, for which the second AH bifurcation occurs at $\tau_{1}^{\textrm{AH2}}$$=$$11.1544$ and is outside the range  of $\tau$ covered by this graph. The first AH bifurcations for each point are marked by filled cyan circles, and the second AH bifurcations by empty circles on the green vertical lines. 

Next, consider non-local bifurcations occurring in this system. 
 As explained in Section~\ref{sec_theory}, we expect at least two kinds of homoclinic bifurcations leading to the disappearance of localized attractors.  These would constitute slightly different versions of the same mechanism, based on global bifurcations, behind the elimination of the barriers between the local minima of $V$, which are embodied in the manifolds separating the basins of attraction of localized attractors. 

The simplest  homoclinic bifurcation occurs when the manifolds of a saddle-focus fixed point at a landscape maximum close to form a homoclinic loop. A more complex scenario takes place if below the homoclinic bifurcation, the saddle-focus at a maximum has undergone AH bifurcation and gave birth to a saddle cycle. Then the homoclinic    ``loop"  is formed by the manifolds of this saddle cycle rather than those of the fixed point.  Note, that the unstable manifold of the saddle cycle has dimension two, rather than one in the case of the simplest saddle fixed point. The structure formed by the manifolds of the saddle cycle at the homoclinic bifurcation is more complex than the relatively simple loop created by the manifolds of the saddle point. For this reason, when describing the homoclinic bifurcation involving the saddle cycle, we refer to a ``loop", rather than the loop.    In either case, the end result is the disappearance of the local attractor as a result of its collision with either the saddle point, or the saddle cycle. 

With this, as predicted by Shilnikov's theorem for a homoclinic loop of a saddle-focus in ODEs \cite{Shilnikov_chaos_from_homoclinic_loop_DANSSSR65,Shilnikov_homoclinic_loop_MUSSR68},  depending on the value of its saddle quantity $\sigma$, before such a collision one can expect two different kinds of dynamics. Namely, $\sigma$$=$$\lambda_1$$+$$Re(\lambda_{2,3})$, where $\lambda_1$ is the positive real eigenvalue of the fixed point, and $\lambda_{2,3}$ are eigenvalues with the negative real parts closest to zero. If $\sigma$$<$$0$, the homoclinic loop is expected to be ``safe" and should coincide with the limit cycle at the instant of collision. This result was verified for a special form of a DDE \cite{Walther_DDE_heteroclinic_to_periodic_TAMS85,Walther_DDE_periodic_orbits_from_homoclinic_BCP89}.  If $\sigma$$>$$0$, the loop is expected to be ``dangerous", and at $\tau$ just below the loop formation the localized attractor should be chaotic. However, to the best of our knowledge, this was not verified for DDEs. Regardless of the kind of the homoclinic loop, the localized attractor should disappear, but in Fig.~\ref{fig_5_well_low5_broad5_bd}(b) we nevertheless indicate the  values of $\tau$ at which 
the saddle quantities of all saddle-foci at the maxima of $V$ change sign from negative to positive by filled yellow circles on the vertical green lines. 

We can see that the type of a homoclinic bifurcation can be predicted from the knowledge of the eigenvalues of the respective saddle point at the instant of bifurcation. However,   the parameter values, at which this or any other non-local  bifurcation occur,  can be detected only through numerical simulations by observing the solutions. Therefore, we reveal all non-local bifurcations in (\ref{dde_whole})  with the landscape (\ref{eq_5well_low5_bro5}) by looking at the phase portraits projected on the plane $($$x(t)$, $x(t-\tau)$$)$.  In Fig.~\ref{fig_5_well_low5_broad5_pp}, panels (a)--(g) illustrate such bifurcations one by one in the order of their occurrence as $\tau$ increases. Each panel shows all fixed points of the system, with black circles indicating the landscape minima $x_i^{\textrm{min}}$, which can be stable or unstable at different $\tau$, and yellow circles indicating always unstable points at the landscape maxima $x_j^{\textrm{max}}$. Note, that all fixed points lie on the diagonal $x(t)$$=$$x(t$$-$$\tau)$.  For the ease of reference, in (a) we label all the landscape maxima 
$x_j^{\textrm{max}}$  as S$j$ with $j$$=$$1,\ldots,4$, and all attractors at or around the minima $x_i^{\textrm{min}}$  as A$i$ with $i$$=$$1,\ldots,5$. 
In each panel, the green line in the left-hand part shows an attractor about to disappear as a result of a bifurcation; the blue line in the right-hand part of the same panel shows an attractor to which the system converges after the bifurcation from the initial conditions where the vanished attractor was. The exceptions here are panels (a)--(b), whose right-hand parts show the whole trajectories converging to the fixed points, rather than the resultant fixed points only.  The respective values of $\tau$ are given in the fields of every phase portrait. Note, that in (a)--(d) there might co-exist oscillatory attractors not involved in the given homoclinic bifurcation, but these are not shown to avoid confusion. 

The first homoclinic bifurcation occurs when the limit cycle A5 (green line in (a)), born at $\tau$$\approx$$0.35409$ around  $x_5^{\textrm{min}}$, is the first to collide with the nearest saddle S4 at $\tau$$\approx$$0.635$ to form a homoclinic loop. As seen from Fig.~\ref{fig_5_well_low5_broad5_bd}(b), this collision occurs before the saddle quantity $\sigma$ turns positive, so the homoclinic loop is ``safe".  After attractor A5 vanishes, the phase trajectory (blue line in (a)) goes to the neighbouring attractor A4, which is the stable fixed point $x_4^{\textrm{min}}$. 

The second non-local bifurcation is associated with a sudden disappearance of a chaotic attractor A3 around $x_3^{\textrm{min}}$ (green line in Fig.~\ref{fig_5_well_low5_broad5_pp}(b)). No collision with any saddle-focus is observed here. Moreover, from Fig.~\ref{fig_5_well_low5_broad5_bd}(b) one can see that no saddle cycles were born from the nearby saddle-foci S2 and S3 before this  global bifurcation,  since both S2 and S3 are below the first AH bifurcation at  $\tau$$=$$1.057$. Thus,
 there could be no collision with a saddle cycle either. We can only hypothesize that some heteroclinic connection might  be involved, in which a manifold of one saddle object connects to another saddle object, but we cannot verify this here. In any case, the second localized attractor disappears as a result of a non-local bifurcation at $\tau$$\approx$$1.0565$, and the phase trajectory (blue line in (b)) goes to the stable fixed point A4. Note, that although $x_3^{\textrm{min}}$ underwent AH bifurcation before $x_5^{\textrm{min}}$ (see Table I), the attractor A3 vanishes only after A5 because its potential well is wider. 

The third non-local bifurcation occurs at $\tau$$\approx$$1.823$ to the limit cycle A1 (green line in (c)) born from $x_1^{\textrm{min}}$ at $\tau$$\approx$$1.0673$. This bifurcation is of the simplest type since the saddle quantity is negative. After the bifurcation, the phase trajectory converges to the closest attractor A2, which is a limit cycle (blue line in (c)). 

Next, at $\tau$$\approx$$1.957$ disappears an attractor A4 (green line in (d)). Note, that at the instant of homoclinic bifurcation, the relevant saddle point S4 at $x_4^{\textrm{max}}$ is above the first AH bifurcation, 
 which implies that there is a saddle cycle around $x_4^{\textrm{max}}$, whose co-dimension-one (infinite-dimensional) stable and two dimensional unstable manifolds form the relevant homoclinic ``loop". The   
attractor about to vanish   in the bifurcation  is chaotic  (compare with simpler cases in \cite{Janson_two-well_delay_2021}). Above this homoclinic bifurcation 
there remains no attractor localized around $x_4^{\textrm{min}}$, and the phase trajectory starting from its vicinity goes to the limit cycle A2 (blue line in (d)). 

The last local attractor to disappear is  A2, which at $\tau$$\approx$$4.067$ collides with a saddle cycle born from the saddle S2 and is therefore chaotic before this collision (green line in (e)). Above this bifurcation, no localized attractors are left. However, a larger attractor embracing all fixed points except $x_1^{\textrm{min}}$ and $x_1^{\textrm{max}}$ is formed (blue line in (e)) presumably due to an intersection of some manifolds or some heteroclinic connection. As $\tau$ grows further, this attractor goes beyond the saddle $x_1^{\textrm{max}}$ (green and blue lines in (f)). A large attractor spanning all fixed points exists for a relatively large range of $\tau$ values, and it remains chaotic for most of this range, as seen from Fig.~\ref{fig_5_well_low5_broad5_pp}(b)). However, at $\tau$$\approx$$7.380588$ the final non-local bifurcation occurs, which destroys the basin of attraction of this large chaos, and the phase trajectory goes to infinity (blue line in (g)).

Note, that the last localized attractor to disappear here is around $x_2^{\textrm{min}}$, which is the lowest and the flattest minimum whose potential well is wide enough (see blue line in Fig.~\ref{fig_5_well_low5_broad5_pp}(a)). The largest flatness (the smallest 
value of $|J|$) of this minimum ensures that the limit cycle is born from it at the largest value of $\tau$ as compared to other minima. The sufficiently large distance between this minimum and the nearest edge of the respective well ensured that for the attractor born from the minimum there is sufficient room to grow in size before  disappearing, and thus to outlive all other localized attractors as $\tau$ grows.

Now, when running the simulation from the initial conditions on the large chaotic attractor at some sufficiently large value of $\tau$$=$$7$, and slowly decrease $\tau$ to zero, as illustrated in Fig.~\ref{fig_5_well_low5_broad5_bd}(c),  the phase trajectory ends up at the lowest minimum $x_2^{\textrm{min}}$ automatically, as desired within an ideal global optimization scenario.

\subsection{Reaching the flattest minimum}
\label{sec_flattest}

\begin{figure}
\includegraphics[width=0.4\textwidth]{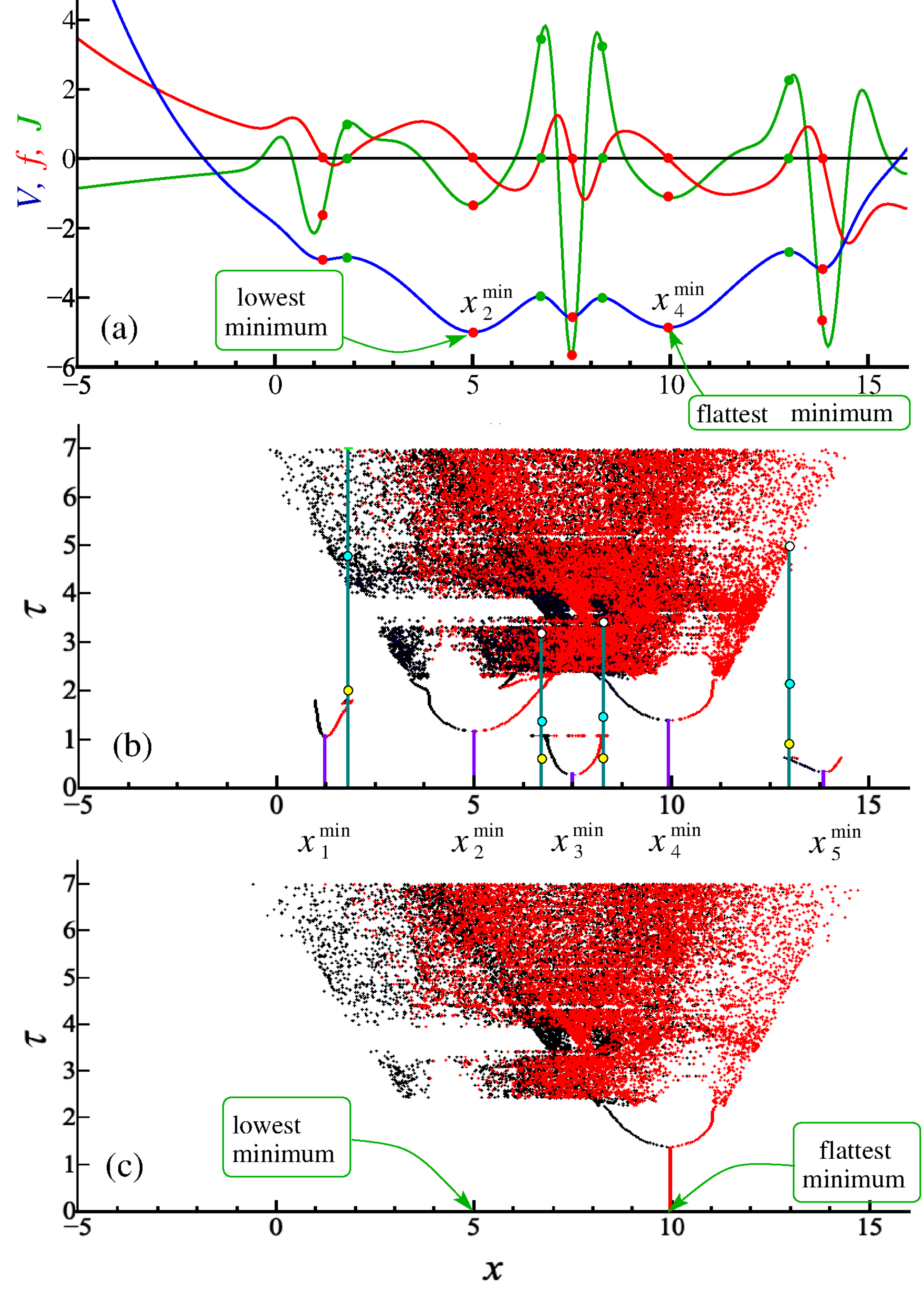}
\caption{Illustration of the case where the lowest minimum $x_2^{\textrm{min}}$ of $V$   is not the flattest, and the flattest minimum is $x_4^{\textrm{min}}$. 
{\bf (a)} Landscape $V$ given  by (\ref{eq_5well_low5_bro10}) (blue line), the function $f$$=$$V'$ of (\ref{dde_whole}) (red line) and its derivative $J$$=$$f'$ (green line). Red/green circles show (i) on blue line, the positions of minima/maxima of $V$, (ii) on red line, the locations of respective fixed points, (iii) on green line, the values of Jacobian $J$ at these fixed points. 
{\bf  (b)} Bifurcation diagram of (\ref{dde_whole})   with landscape  (\ref{eq_5well_low5_bro10}) as $\tau$ grows, notations are as in Fig.~\ref{fig_5_well_low5_broad5_bd}(b). 
{\bf  (c)} Behavior of (\ref{dde_whole})   with $V$ given by  (\ref{eq_5well_low5_bro10}) as $\tau$ {\it decreases} slowly from $\tau$$=$$7$ starting from a point on the large chaotic attractor. Local minima (black dots) and maxima (red dots) of the solution are shown. The system automatically converges to the flattest minimum $x_4^{\textrm{min}}$, which is slightly higher than the global minimum  $x_2^{\textrm{min}}$ (see (a) and Table I under (\ref{eq_5well_low5_bro10})).  }
\label{fig_5_well_low5_broad10_bd}
\end{figure}

\begin{figure*}
\includegraphics[width=0.8\textwidth]{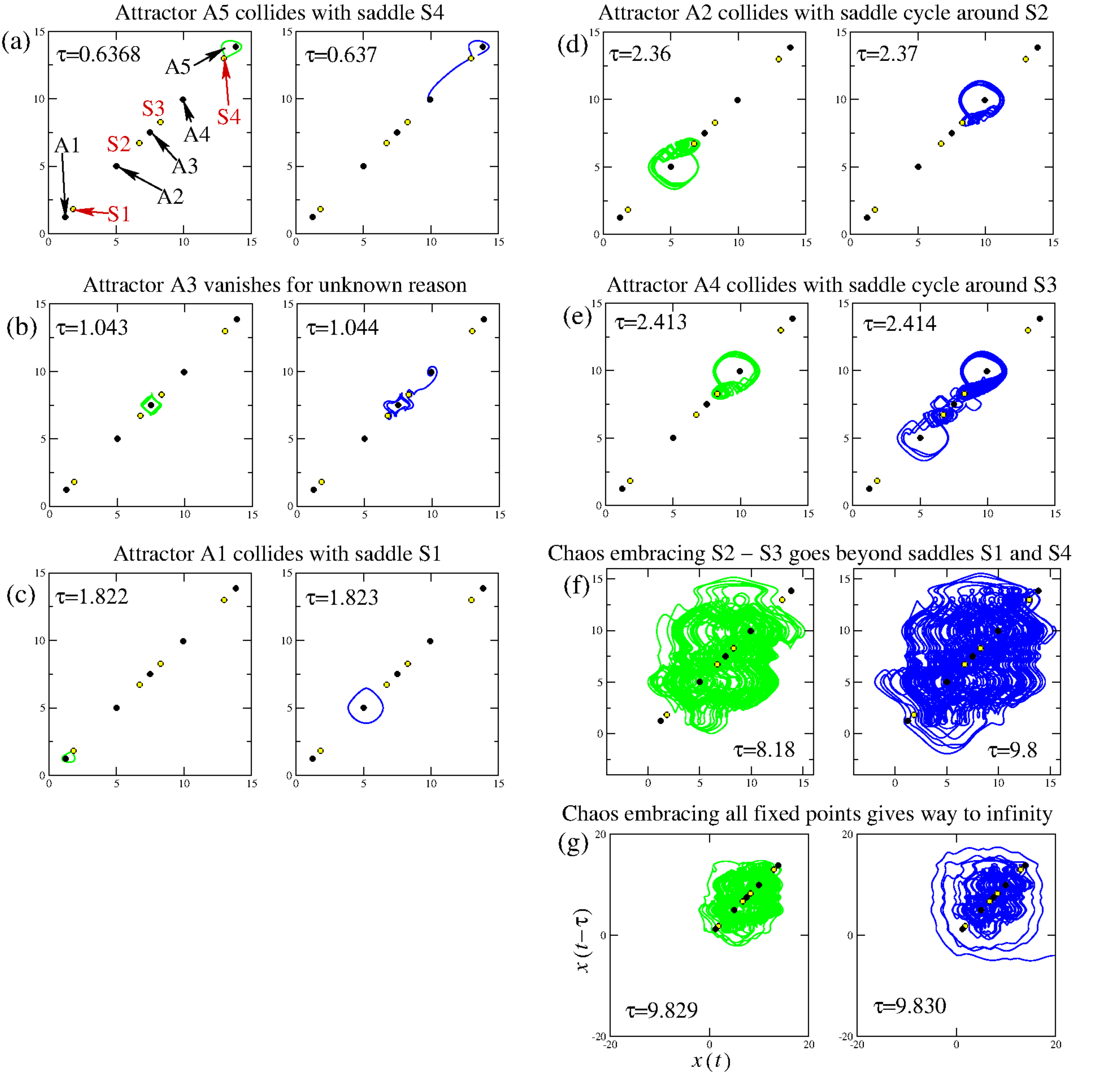}
\caption{
Phase portraits on the plane $($$x(t)$, $x(t-\tau)$$)$ illustrating a sequence of non-local bifurcations in (\ref{dde_whole})   with landscape  (\ref{eq_5well_low5_bro10}) as $\tau$ increases. 
Here, the lowest minimum of $V$ is $x_2^{\textrm{min}}$, and the flattest one is $x_4^{\textrm{min}}$. Fixed points at minima/maxima of $V$ are given by black/yellow circles. 
Panel (a)  shows 
labels of attractors A1 -- A5 at or around the minima, and of saddle fixed points S1 -- S4 at the maxima,  as they are referred to in text. 
The value of  $\tau$ increases from  (a) to (g).   Each panel illustrates the situation just before (left) and immediately after (right)  a certain non-local bifurcation with the values of $\tau$ given in the fields of each plot. \\
{\bf (a)} Before (left) and after (right) the safe homoclinic loop of the saddle point S4 (similar to Fig.~\ref{fig_5_well_low5_broad5_pp}(a)). Green line: limit cycle A5  before bifurcation. Blue line: phase trajectory starting near  A5 and converging to the minimum A4  after bifurcation.\\
{\bf (b)} Before (left) and after (right) the global bifurcation eliminating attractor  A3 (similar to Fig.~\ref{fig_5_well_low5_broad5_pp}(b)). Green line: chaotic attractor  A3 before bifurcation. Blue line: phase trajectory starting near  A3 and converging to the minimum A4  after bifurcation.  \\
{\bf (c)} Before (left) and after (right) the safe homoclinic loop of the saddle point  S1 (similar to Fig.~\ref{fig_5_well_low5_broad5_pp}(c)). Green line: limit cycle A1 before bifurcation. Blue line: limit cycle A2 to which trajectory converges  after bifurcation starting from the vicinity of  A1.  \\
{\bf (d)} Before (left) and after (right) the homoclinic ``loop" formed by the manifolds of the saddle cycle around  S2 (different from Fig.~\ref{fig_5_well_low5_broad5_pp}(d)). Green line: chaos A2 before bifurcation. Blue line: chaos A4 attracting the trajectory  after bifurcation  from the vicinity of  A2. \\
{\bf (e)} Before (left) and after (right) the homoclinic ``loop" formed by the manifolds of the saddle cycle around  S3 (different from Fig.~\ref{fig_5_well_low5_broad5_pp}(e)). Green line: chaotic attractor  A4 before bifurcation. Blue line: non-localized chaotic attractor born after the bifurcation. \\
{\bf (f)} Before (left) and after (right) the global bifurcation giving birth to large chaos embracing all fixed points (similar to Fig.~\ref{fig_5_well_low5_broad5_pp}(f)). Green line: smaller chaotic attractor  before bifurcation. Blue line: large chaotic attractor born from the bifurcation. \\
{\bf (g)} Death of the last attractor (similar to Fig.~\ref{fig_5_well_low5_broad5_pp}(g)). Green line: large chaos before bifurcation. Blue line: trajectory going to infinity after bifurcation.  
}
\label{fig_5_well_low5_broad10_pp}
\end{figure*}

 In the example considered here, $V$ is designed in order to achieve a different order of  deaths of the localized attractors as $\tau$ increases, comparing to  (\ref{eq_5well_low5_bro5}), by appropriately adjusting the flatnesses, depths and locations of the relevant minima, as described below.  The landscape $V(x)$ is specified by Eq.~(\ref{eq_5well_low5_bro10}) and given by blue line in Fig.~\ref{fig_5_well_low5_broad10_bd}(a).  Similarly to the landscape described by  (\ref{eq_5well_low5_bro5}) and 
considered in Sec.~\ref{sec_global}, 
  here there are 
five 
 minima.  Four of these minima, namely $x_1^{\textrm{min}}$  to $x_4^{\textrm{min}}$, are at approximately the same positions, and their respective potential wells are of approximately the same width as in (\ref{eq_5well_low5_bro5}). Inside each well of $V$, the function $f$ of (\ref{dde_whole}) (red line in Fig.~\ref{fig_5_well_low5_broad10_bd}(a)) does not oscillate, so the shape of $V$ is relatively simple. 

The global minimum is still $x_2^{\textrm{min}}$, however,  it  is 
no longer
the flattest of all minima. 
The minimum $x_4^{\textrm{min}}$ is slightly higher than 
$x_2^{\textrm{min}}$ and also slightly flatter, i.e. its $|J|$  is slightly smaller (see Table I below Eq.~(\ref{eq_5well_low5_bro10})). 
With this, because $x_5^{\textrm{min}}$ is now located further to the right than in (\ref{eq_5well_low5_bro10}), the distances from both 
$x_2^{\textrm{min}}$ and $x_4^{\textrm{min}}$ to their respective nearest well edges are very close. 

Figure~\ref{fig_5_well_low5_broad10_bd} has the same structure and notations as Fig.~\ref{fig_5_well_low5_broad5_bd}. One can see that the bifurcation diagram in (b) is qualitatively similar to the one of Fig.~\ref{fig_5_well_low5_broad5_bd}(b), and all the same bifurcations are taking place here as $\tau$ increases, albeit in a slightly different order and at different values of $\tau$. This is also evident from comparing Figs.~\ref{fig_5_well_low5_broad5_pp} and \ref{fig_5_well_low5_broad10_pp} where the non-local bifurcations are illustrated with the phase portraits.  

The most important distinction from the case of Sec.~\ref{sec_global} is that the last localized attractor surviving at the higest value of $\tau$ is chaos A4 around $x_4^{\textrm{min}}$. In Fig.~\ref{fig_5_well_low5_broad10_pp}(e) green line in the left part shows this attractor just before the homoclinic bifurcation at $\tau$$=$$2.413$, and blue line in the right part shows the attractor to which the phase trajectory converges after the homoclinic bifurcation at $\tau$$=$$2.414$, which is chaos spanning three minima and two maxima. The homoclinic bifurcation occurs at $\tau$$\approx$$2.4135$ and consists of the closure of manifolds of the saddle cycle around   
$x_3^{\textrm{max}}$ (S3). 

After the disappearance of all localized attractors, the further increase of $\tau$ leads to the further growth of the single large chaotic attractor, until it goes beyond the outermost maxima as illustrated in Fig.~\ref{fig_5_well_low5_broad10_pp}(g) and ends up spanning all fixed points. With this, the growth of this single remaining attractor is limited by the final non-local bifurcation at $\tau$$\approx$$9.8295$, after which the phase trajectory tends to infinity (Fig.~\ref{fig_5_well_low5_broad10_pp}(f)).

If we launch the system from the initial conditions on the large chaotic attractor at $\tau$$=$$7$, and its behavior is observed as $\tau$ slowly decreases to zero, the system converges to the non-global minimum $x_4^{\textrm{min}}$, which is the flattest minimum, as illustrated in Fig.~\ref{fig_5_well_low5_broad10_bd}(c).

\subsection{Reaching an arbitrary minimum}

\label{sec_arbitrary}

Here we construct an example of $V$ with the same number of minima as in previously considered cases (\ref{eq_5well_low5_bro5}) and (\ref{eq_5well_low5_bro10}), but with a more intricate shape of some wells.  By doing so, we assess whether the enhanced non-linearity of the system is likely to disrupt the hypothesized  and provisionally verified chain of delay-induced global bifurcations destroying local attractors one by one, leading to  global chaos at large values of $\tau$, and thus removing the barriers that without the delay prevented the phase trajectory from approaching some (most) minima of $V$. 

\begin{figure}
\includegraphics[width=0.4\textwidth]{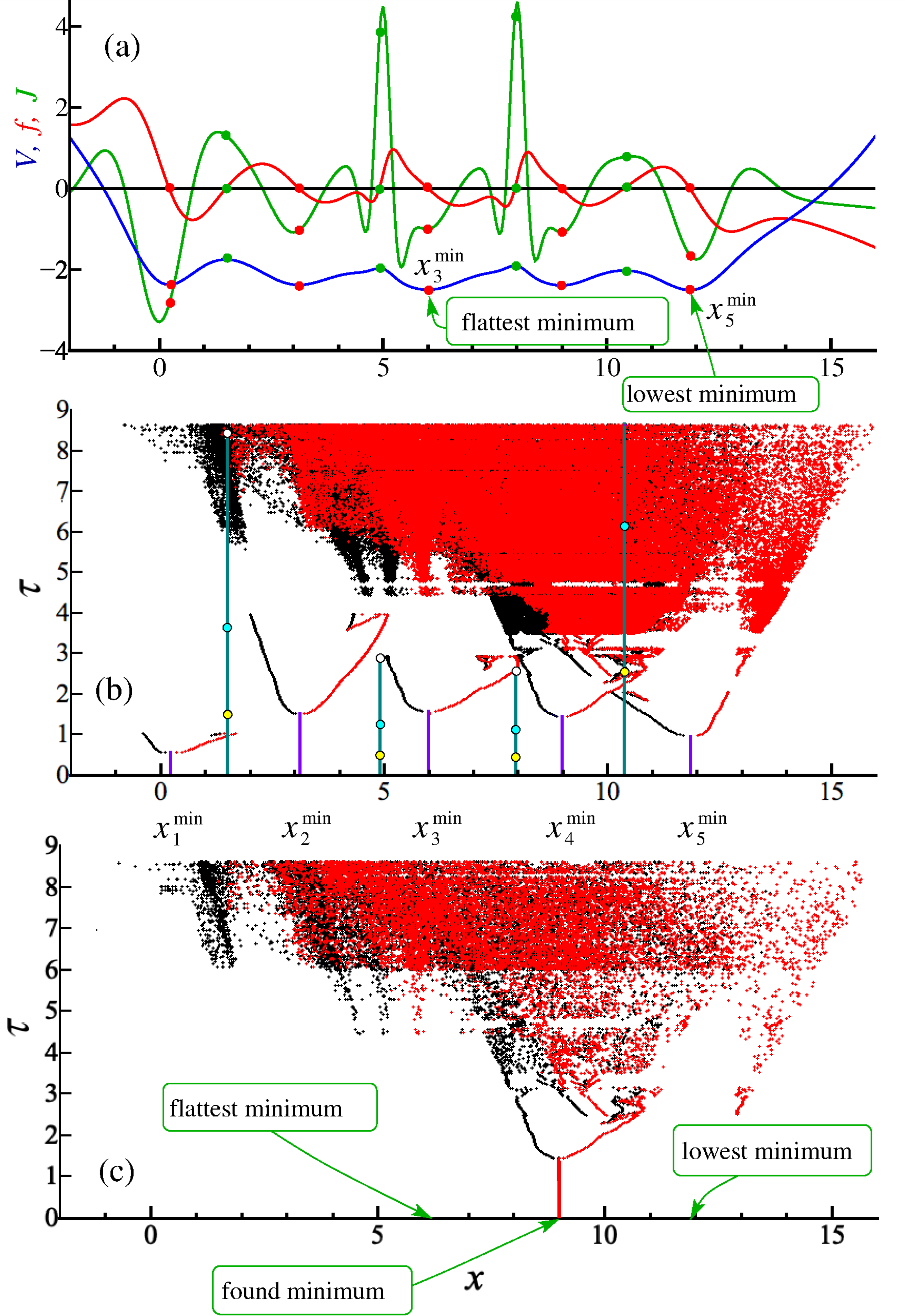}
\caption{Illustration of the case where the landscape $V$ has complex shape, i.e. $f$$=$$V'$ oscillates between consecutive  zero crossings. Here the lowest minimum of $V$ is $x_5^{\textrm{min}}$, the flattest one is $x_3^{\textrm{min}}$, and the found  one is $x_4^{\textrm{min}}$. {\bf (a)} Landscape $V$ given  by (\ref{eq_5well_chaos_mess})  (blue line), the function $f$$=$$V'$ of (\ref{dde_whole}) (red line) and its derivative $J$$=$$f'$ (green line). Red/green circles show (i) on blue line, the positions of minima/maxima of $V$, (ii) on red line, the locations of respective fixed points, (iii) on green line, the values of Jacobian $J$ at these fixed points. 
{\bf  (b)} Bifurcation diagram of (\ref{dde_whole})   with landscape  (\ref{eq_5well_chaos_mess}) as $\tau$ grows, notations are as in Fig.~\ref{fig_5_well_low5_broad5_bd}(b). 
{\bf  (c)} Behavior of (\ref{dde_whole})   with $V$ given by   (\ref{eq_5well_chaos_mess})  as $\tau$ {\it decreases} slowly from $\tau$$=$$8.6$ starting from a point on the large chaotic attractor. Local minima (black dots) and maxima (red dots) of the solution are shown. The system automatically converges to the flattest minimum $x_4^{\textrm{min}}$, which is slightly higher than the global minimum  $x_5^{\textrm{min}}$ and slightly more concave than the flattest minimum $x_3^{\textrm{min}}$ (see (a) and Table I under (\ref{eq_5well_chaos_mess})). 
  }
\label{fig_5_well_chaos_mess_bd}
\end{figure}

We are also interested in testing the proposed approach to optimization in the difficult situation where the deepest wells are only subtly different in their depths. The landscape $V$  meeting the above requirements is specified by    
(\ref{eq_5well_chaos_mess}) in Table~I and shown by blue line in Fig.~\ref{fig_5_well_chaos_mess_bd}(a). Here, just like in the previous two cases considered, $V$ has five minima. However, in the range $x$$\in$$ [3,10]$ the 
function $f$$=$$V'$ demonstrates small oscillations between consecutive pairs of zero-crossings,  which make the respective DDE considerably more nonlinear.  The depths of the five local minima are only slightly different from each other (see Table~I below Eq.~(\ref{eq_5well_chaos_mess})), and the global minimum $x_5^{\textrm{min}}$ is only slightly lower than the other minima. 

As $\tau$ grows, all five fixed points $x_1^{\textrm{min}}$--$x_5^{\textrm{min}}$ undergo AH bifurcations and give birth to stable limit cycles, which grow in size with $\tau$ in full analogy with the previous cases considered, as illustrated in Fig.~\ref{fig_5_well_chaos_mess_bd}(b).

However, non-local bifurcations occur in a less predictable manner than with previous examples. In what follows, we indicate  similarities and distinctions from the cases considered above in the scenarios developing as $\tau$ grows. 

\begin{figure*}
\includegraphics[width=0.8\textwidth]{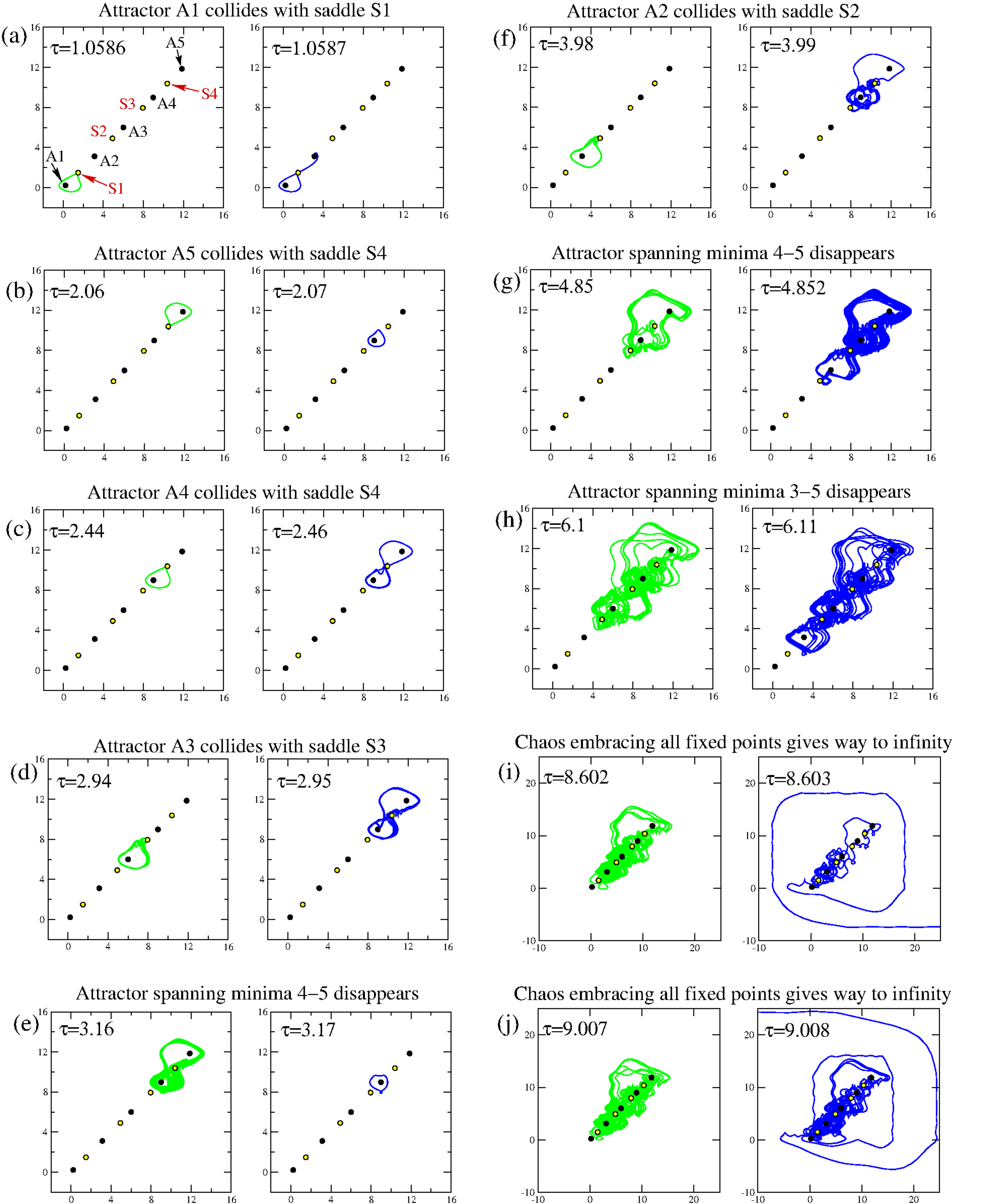}
\caption{
Phase portraits on the plane $($$x(t)$, $x(t-\tau)$$)$ illustrating a sequence of non-local bifurcations in (\ref{dde_whole})  with landscape  (\ref{eq_5well_chaos_mess}) as $\tau$ increases. Here, $V$ has more intricate shape than in cases illustrated by Figs.~\ref{fig_5_well_low5_broad5_pp} and \ref{fig_5_well_low5_broad10_pp}.
Minimum  $x_5^{\textrm{min}}$ is the lowest,  $x_3^{\textrm{min}}$ is the flattest, and  $x_4^{\textrm{min}}$ is found with optimization procedure,  but their depths are very similar. Fixed points at minima/maxima of $V$ are given by black/yellow circles. 
Panel (a)  shows  
labels of attractors A1 -- A5 at or around the minima, and of saddle fixed points S1 -- S4 at the maxima,  as they are referred to in text. 
Delay  $\tau$ increases from  (a) to (j), with the values of $\tau$ given in the fields of each plot.\\  
{\bf (a)/(b)} Before (left) and after (right) the safe homoclinic loop of the saddle point at the maximum S1/S4. Green line: limit cycle A1/A5 before bifurcation. Blue line: phase trajectory starting near  A1/A5 and converging to fixed point A2/ limit cycle A4  after bifurcation. \\
{\bf (c)} Before (left) and after (right) the safe homoclinic loop of the saddle point  S4. Green line: limit cycle A4 before bifurcation. Blue line: non-localized attractor attracting trajectory  from the vicinity of  A4.  \\
{\bf (d)} Before (left) and after (right) the homoclinic ``loop" formed by the manifolds of the saddle cycle around  S3 (similar to Fig.~\ref{fig_5_well_low5_broad10_pp}(e)). Green line: chaos A3 before bifurcation. Blue line: non-localized chaos attracting the trajectory  after bifurcation  from the vicinity of  A3. \\
{\bf (e)} Temporary disappearance of the first non-localized attractor (see gap in  Fig.~\ref{fig_5_well_chaos_mess_bd}(b)). Green line: non-localized chaos before bifurcation. Blue line: limit cycle A4 attracting the trajectory from where chaos was at $\tau$$=$$3.16$. \\
{\bf (f)} Before (left) and after (right) the homoclinic ``loop" formed by the manifolds of the saddle cycle around  S2.   Green line:  last localized attractor A2. Blue line: reappeared non-localized chaos (compare with (d)). \\
{\bf (g)/(h)} Smaller non-localized attractors replaced by bigger ones. Smaller chaos before (green line) and  larger chaos after (blue line) bifurcation.  \\
{\bf (i)} Sensitivity to initial conditions and parameters. Green line: large chaos. Blue line: trajectory goes to infinity after a tiny change in $\tau$.  \\
{\bf (j)} Death of the last attractor. Green line: large chaos before bifurcation. Blue line: trajectory goes to infinity after bifurcation.  
}
\label{fig_5_well_chaos_mess_pp}
\end{figure*}

We start from listing the \emph{similarities}. Firstly, all five localized attractors disappear through the homoclinic bifurcations in the same manner as before, as illustrated in 
Figs.~\ref{fig_5_well_chaos_mess_pp}(a)--(d) and (f). With this, the nature of the homoclinic bifurcation is determined by whether the relevant saddle-focus fixed point was before or after the first AH bifurcation. Namely, in (a)--(c) the limit cycles born from $x_1^{\textrm{min}}$,  $x_5^{\textrm{min}}$ and $x_4^{\textrm{min}}$ collide with the saddle-foci $x_1^{\textrm{max}}$ and $x_4^{\textrm{max}}$ and form safe homoclinic loops, because the latters are not only below their first AH bifurcations, but also below  the instants where their saddle quantities $\sigma$ become positive. On the other hand,  (d) and (f) illustrate homoclinic bifurcations representing the closures of manifolds of 
saddle cycles born from $x_3^{\textrm{max}}$ and $x_2^{\textrm{max}}$, respectively.

Note, that in the  given landscape $V$, $x_2^{\textrm{max}}$ and $x_3^{\textrm{max}}$ are much sharper than in (\ref{eq_5well_low5_bro5}) and (\ref{eq_5well_low5_bro10}), as can be seen from the high splashes of function $J$$=$$V''$ in Fig.~\ref{fig_5_well_chaos_mess_bd}(a) (green line). The high values of $J$ lead to AH bifurcations of these fixed points occurring at much smaller values of $\tau$ than in both previous examples. However, the homoclinic bifurcations involving manifolds of the saddle cycles around these maxima occur in the same manner as in the previous two examples (compare Figs.~\ref{fig_5_well_chaos_mess_pp}(d), (f) with e.g. Figs.~\ref{fig_5_well_low5_broad10_pp}(d)--(e)), although at the instants of homoclinic collisions both $x_2^{\textrm{max}}$ and $x_3^{\textrm{max}}$  are already above the second AH bifurcation. 

The second similarity is that at a sufficiently large $\tau$, the system possesses a single large chaotic attractor embracing all fixed points (compare the left parts of Figs.~\ref{fig_5_well_chaos_mess_pp}(i)--(j) with those of  Figs.~\ref{fig_5_well_low5_broad5_pp}(g) and \ref{fig_5_well_low5_broad10_pp}(g)). Thirdly, at even larger $\tau$, this large chaos disappears too, and the system goes to infinity 
(compare the right part of Fig.~\ref{fig_5_well_chaos_mess_pp}(j) with those of Figs.~\ref{fig_5_well_low5_broad5_pp}(g) and \ref{fig_5_well_low5_broad10_pp}(g)). Finally and perhaps most importantly, as could be expected from the previous argument and in agreement with previous cases, the localized attractor surviving at the largest value of $\tau$ is A2 whose potential well is the flattest and of a similar width with other wells (see Fig.~\ref{fig_5_well_chaos_mess_pp} (f)).

Next, we describe the \emph{distinctions} from the relatively predictable scenarios involving relatively simple shapes of $V$ given by (\ref{eq_5well_low5_bro5}) and (\ref{eq_5well_low5_bro10}). Firstly, unlike in the case above, here attractors involving two or more minima can coexist with attractors localized around a single minimum. Namely, in  Figs.~\ref{fig_5_well_chaos_mess_pp}(c)--(f) one can see a chaotic attractor spanning  minima $x_4^{\textrm{max}}$ and $x_5^{\textrm{max}}$, which exists while the localized attractors A2, A3 and A4 are not yet destroyed via homoclinic bifurcations. 

Secondly, as seen from Figs.~\ref{fig_5_well_chaos_mess_bd}(b) and \ref{fig_5_well_chaos_mess_pp}, in homoclinic bifurcations involving $x_2^{\textrm{max}}$--$x_4^{\textrm{max}}$, counterintuitively, the localized attractor collides with the saddle object (point or cycle), which is not the closest to the minimum from which this attractor had originated. For example,  Fig.~\ref{fig_5_well_chaos_mess_pp}(d) shows an attractor A3 around $x_3^{\textrm{min}}$ undergoing a homoclinic bifurcation with the saddle cycle around S3 at $x_3^{\textrm{max}}$, although $x_2^{\textrm{max}}$ and the saddle cycle around it are both located closer.

Thirdly, and most significantly, the minimum attracting the system while $\tau$ slowly decreases from the  value of $8.6$ is not the one which survives at the largest value of $\tau$, i.e. not $x_2^{\textrm{min}}$, as can be seen from  Fig.~\ref{fig_5_well_chaos_mess_bd}(c). Instead, it is the $x_4^{\textrm{min}}$, whose homoclinic bifrucation with the saddle-focus S4 at $x_4^{\textrm{max}}$ gave rise to the chaotic attractor embracing $x_4^{\textrm{min}}$ and $x_5^{\textrm{min}}$ (see blue line in Fig.~\ref{fig_5_well_chaos_mess_pp}(c)). The reason is that the large chaos shown in the left part of (i) develops from the latter attractor as  $\tau$ grows. When at 
$\tau$$=$$8.6$  the phase trajectory is launched from the only attractor available, as    $\tau$ decreases, this chaotic attractor undergoes a cascade of non-local bifurcations, as a result of which it becomes confined within the smaller and smaller number of minima. As a result, the phase trajectory does not go beyond $x_3^{\textrm{max}}$ (S3) since it settles down on the attractor A4 (see green line in (c)), which exists for all values of $\tau$ up to zero. 

Interestingly, at  $\tau$ above $8.6$ the basin of attraction of large chaos seems to become very close to the attractor itself, and is apparently quite riddled, since a very small disturbance of initial conditions can lead to the phase trajectory going to infinity even before  the attractor disappears completely at $\tau$$=$$9.008$, as illustrated in Figs.~\ref{fig_5_well_chaos_mess_pp}(i). 

The attractors spanning more than one minimum appear here despite the existence of localized attractors.  By analogy with similar phenomena in ODEs \cite{Li_Moon_heteroclinic_chaos_JSV90}, it is reasonable to suggest that they could be born as a result of repeated formation of homoclinic loops and heteroclinic connections. Just like in the previous simpler cases, here there is an abundance of saddle objects with manifolds attached to them, which can potentially form loops and heteroclinic connections. However, due to the complex shape of $f$, the   probability of a homoclinic or a heteroclinic bifurcation at any given value of $\tau$  seems to be higher here. 

The current example, deliberately constructed as a difficult case, is an excellent illustration that in nonlinear systems with delay, it is generally impossible to predict, before performing numerical bifurcation analysis,  the behavior as the delay varies.  However, it is remarkable that in its most essential part, the predicted scenario is realized even despite the enhanced complexity of the nonlinear function in the DDE  (\ref{dde_whole}), provided that $V$ is a smooth multi-well function going to infinity outside the domain containing the extrema.

\subsection{Summary of findings}
\label{summary}

Here we summarize the results reported in the three examples considered in Sections~\ref{sec_global}, \ref{sec_flattest} and \ref{sec_arbitrary},  which illustrate different mechanisms of barrier-breaking between the local minima.  We reiterate that generally it is impossible to predict  without numerical analysis the behavior of nonlinear systems with delay as the delay changes, as illustrated  particularly well by the case of Section~\ref{sec_arbitrary}. However, for delay systems in a special form (\ref{dde_whole}) with $V(x)$ being a multi-well landscape tending to positive infinity as $|x|$ goes to infinity,  it appears possible to make certain  qualitative predictions with regard to the bifurcations occurring as the delay $\tau$ increases. Firstly, as has been well known for general delay equations, one can accurately predict the values of $\tau$, at which local Andronov-Hopf bifurcations occur for all fixed points.
Also, in the systems of the form (\ref{dde_whole}) with single-well and/or single-hump $V$ 
it is possible to predict the occurrence of a non-local homoclinic bifurcation with the growth of $\tau$ to some extent, 
although the exact value of $\tau$ cannot be predicted and the exact form of this bifurcation can be determined only when the respective $\tau$ is known at least approximately.  For systems with multi-well landscapes, we envisioned  
that   localized attractors growing in size with $\tau$ would eventually collide with saddle objects and their manifolds in non-local bifurcations. 

However, although non-local bifurcations are quantitatively unpredictable, some   qualitative universal phenomena induced by such bifurcations  have been hypothesized and 
revealed 
in all three cases considered. Note, that that all attractors grow in size with the delay until they experience some non-local bifurcation and vanish. Thus,  the first effect is the disappearance of all attractors localized around the minima of $V$ as the delay increases, with the typical reason behind their disappearance being a homoclinic bifurcation involving the manifolds of a nearby saddle-focus fixed point or a saddle cycle. The second phenomenon is the existence of a single attractor spanning all minima of $V$ at sufficiently large values of $\tau$, which  is chaotic for most $\tau$ at which it exists. The third phenomenon is disappearance of all attractors as $\tau$ exceeds a certain threshold, which is different in different systems. 

Besides the coexistence of  attractors confined to different wells of $V$, a typical phenomenon is the coexistence of attractors within the same well. Some of these cases were illustrated  for single- and two-well cases in \cite{Janson_two-well_delay_2021}, but we are not focussing on them specially here because they do not play a significant role in the system behavior when $\tau$ is decreased slowly.  Also, the coexistence of attractors spanning more than one well is possible, particularly with complex shapes of $V$,  such as the one considered in Section~\ref{sec_arbitrary}.

Generally, in the class of delay systems considered here, besides Andronov-Hopf bifurcation, the key role is played by the numerous manifolds of the saddle fixed points and saddle cycles, whose tangencies and closures induce dramatic changes in the system behavior. 
 
In addition to possessing five local minima, the three cases of the landscape function $V(x)$ considered here have one more feature in common, namely, as $x$ tends to $\pm \infty $, they asymptotically tend to the function $k(x-a)^4$, where $k$$>$$0$ and $a$ are some real constants. However,  the universality of the phenomena presented here is confirmed by our consideration of several other forms of the landscape function $V$ with different asymptotic behaviors, which we do not report in this paper  because of the limited space. 

Note, that the original idea of the given research has been to investigate the applicability of delay-induced bifurcations to global optimization. 
The hypothesized bifurcation scenario was partly verified and partly clarified here. With this, it appears that the delivery of the global  minimum with this approach is possible if the shape of the cost function $V$ in the vicinity of this minimum satisfies certain conditions, and if the shapes of all potential wells of $V$ are not too complex. 
Namely, 
the first two cases demonstrate that in the absence of oscillations of $V'$ within individual potential wells of $V$, the slow decrease of $\tau$ from a large value to zero makes the system converge to the minimum, which is the flattest and is far enough from the nearest maximum. This would be  the minimum around which a localized attractor survives at the largest value of $\tau$. If such a minimum is  global, then the proposed %technique 
 procedure will deliver the global minimum. 

\section{Discussion}

\label{sec_discussion}

To summarize, we conceived a 
 phenomenological mathematical  model of a non-quantum  
\emph{analogue} optimizer based on a previously unknown principle behind overcoming the barriers between the minima of the cost function, namely, global bifurcations caused by time delay. We proved the principle by numerical simulations of a scalar version of the model (\ref{GDS_delay}), and explored the possible theoretical limitations of this approach.  As a key prerequisite to our approach, based on a highly limited number of relevant rigorous theoretical results available for non-linear delay equations,  we formulated a hypothesis about the inevitability of a chain of global attractor- and barrier-destroying bifurcations caused by the increase of delay.    Further research would be needed in order to test this approach with non-scalar versions of  (\ref{GDS_delay})  and with more realistic tasks to solve. If its workability is confirmed, it would be interesting to implement the same principle with analogue electronic circuits, which could potentially be more efficient than digital computers in such specialized tasks.  

Like in many earlier works, while developing an approach for global optimization, we introduced an extension of the famous gradient descent method described by Eq.~(\ref{GDS_nodelay}), in which the role of the landscape $V$ is played by the cost function. However, unlike in previous works, which used stochastic perturbations of the relevant evolution equation and employed probabilistic approaches, our trial approach described by (\ref{GDS_delay}) is entirely deterministic and extremely simple in its setting. 
Namely, we delay the   argument of the right-hand side of the ``gradient descent" differential equation by a certain amount $\tau$, which becomes the only control parameter in the system. 

We assume that the optimization technique should start from launching the system (\ref{GDS_delay}) from arbitrary initial conditions at zero $\tau$ and waiting until it reaches one of the local minima, then increasing $\tau$ slowly until we observe the birth of a large attractor and its subsequent disappearance, and noting the respective value of $\tau$. The procedure should be repeated until $\tau$ reaches the value just below the disappearance of the global attractor. After that, in an ideal scenario, a slow decrease of $\tau$ to zero would automatically bring the system to the global minimum of the cost function. Here we revealed the conditions under which this scenario could be possible. 

Just like the simulated and quantum annealings, our 
 approach 
involves overcoming the potential barriers separating local minima of the cost function.  
 However, it  
uses a different principle to achieve this. Namely, in simulated annealing the fictitious massless particle is pushed over the maxima, in quantum annealing the barriers can be in addition penetrated using tunnelling effect \cite{Denchev_quantum_annealing_shape_depenence_PRX16}, and in our delay-based method the barriers are effectively removed via global bifurcations. 

Specifically, we revealed that in a scalar version (\ref{dde_whole}) of (\ref{GDS_delay}) the key phenomenon is homoclinic bifurcation, which occurs repeatedly as the delay increases. The role of the homoclinic bifurcation is to eliminate, one by one, all attractors localized around the minima of the landscape  as the delay grows, and to ultimately give rise to a single chaotic attractor spanning all local minima at some moderately large value of the delay. 

The homoclinic bifurcation in the system considered comes in at least two varieties depending on whether the relevant saddle point located at the maximum of the landscape has undergone Andronov-Hopf bifurcation or not. 
However, in the context of global optimization, the exact form of a particular homoclinic bifurcation is not essential since, regardless of its form, it results in the breakdown of the localized attractor. With this, the localized attractors can disappear through  mechanisms not immediately associable with the homoclinic bifurcations,  for example 
in Figs.~\ref{fig_5_well_low5_broad5_pp}(b) and \ref{fig_5_well_low5_broad10_pp}(b). 

Based on the results of our analysis of (\ref{dde_whole}) with several smooth landscapes $V$, which were constructed to specification in order to assess various 
 versions of the barrier-breaking mechanism operating in  various orders,
we conclude that the delay-induced chain of homoclinic and other global bifurcations destroying local attractors is a 
universal scenario, whose details depend on the shape of the particular function $V$, but whose existence is independent of the latter. 

Standard optimization tools, including simulated annealing, require algorithmic decision-making at numerous stages of the procedure and thus rely on the provision of a digital computer.  It is certainly possible to simulate a   continuous-time system with delay on a computer,   as we do here.   However, we note that the proposed approach  exploits \emph{spontaneous} behavior of  a  dynamical system, which would emerge in an underlying physical device thanks to its inner structure, just like oscillations of a pendulum  in a ``grandfather clock". This approach does not require continual decision-making dependent on the previous step, and hence does not need to rely on a digital computer.  
 It could 
 be implemented in a non-quantum analogue device with a delay, which could be designed to have a high evolution rate and hence achieve a solution reasonably quickly.

By analogy with  optimization techniques based on stochastic extensions of gradient descent, 
it is impossible to guarantee the convergence of (\ref{GDS_delay}) to the global minimum of $V$ by slowly decreasing $\tau$ to zero due to the general impossibility to predict the behavior of nonlinear systems. 
However, if the global minimum is also the flattest and the respective well is wide enough, 
as $\tau$ grows, the attractor around it is likely to be destroyed last. If the wells of the cost function are not too complex in their shape, 
as $\tau$ decreases from the positive value, at which global chaos exists, to zero, the system is likely to end up at the global minimum. 

Note, that the above conditions are quite consistent with those for quantum computers, in which the possibility to achieve a \emph{global} minimum depends on the shape  of the energy landscape \cite{Denchev_quantum_annealing_shape_depenence_PRX16}. 
With this, 
 the  implementability of the proposed principle with analogue electronic devices could  
be an advantage over quantum computers, which although expected to be the best optimizers in theory,  are  difficult  and expensive to make in practice. 
Therefore, this approach deserves to be  explored further as an interesting alternative to 
the available techniques for global optimization.

Obviously, it will be very important to assess the efficiency of the proposed approach by applying it to more realistic challenging problems, and by comparing its performance with that of the standard optimization methods, which will be the subject of future work.  

In addition to probing a novel approach to optimization, our paper contributes to the theory of nonlinear delay differential equations (DDEs) by considering their special class and revealing a universal scenario that unfolds as the delay is increased. Despite the high nonlinearity of the DDEs considered, we show that it is possible to make some qualitative predictions about their behavior. 

In order to improve our understanding of the delay-induced effects in the nonlinear systems of the given class, it would be desirable to verify some of the qualitative predictions and numerical observations made here with more rigorous approaches. For example, it would be important to rigorously address the seeming absence of any attractors at sufficiently large values of delay, or the existence of a single global attractor embracing all fixed points at some moderate delays. Lyapunov functionals could be promising for this purpose, although identification of suitable functionals is not straightforward here.  Addressing these questions would be an interesting task for the future.    

\section{Authors' contributions}

NBJ proposed the research, analysed systems illustrated in the paper, interpreted the results, and wrote the paper.   CJM carried out preliminary testing of the initial hypothesis with examples not illustrated in the paper, and edited the paper. Both authors contributed to the design of research. 

\section{Data availability}

Data sharing is not applicable to this article as no new data were created or analyzed in this study.

\section{Acknowledgements}

The authors are grateful to Alexander Balanov for help in creating software to solve DDEs numerically and for helpful feedback on the draft of the paper, and to Alexander Zagoskin for the discussions of quantum annealing. CJM was supported by EPSRC (UK) during his PhD studies in Loughborough University,  grant EP/P504236/1.

\end{document}